\documentclass[
  prd,preprint,longbibliography,
  showpacs,showkeys,lengthcheck,
  nofootinbib,tightenlines,onecolumn,notitlepage,
  preprintnumbers,superscriptaddress
]{revtex4-1}

\usepackage[utf8]{inputenc}
\usepackage{newtxtext,newtxmath}
\usepackage[mathcal]{euscript}

\usepackage{graphicx}
\usepackage[usenames,dvipsnames]{xcolor}
\graphicspath{{figures/}}
\usepackage{tikz-feynman}

\usepackage{hyperref}
\hypersetup{colorlinks=true,citecolor=blue,linkcolor=blue,urlcolor=blue}
\usepackage{orcidlink}

\usepackage{amsmath}
\usepackage{slashed}
\usepackage{bm}
\usepackage{bbm}
\usepackage{mathtools}

\usepackage[math]{cellspace}
\usepackage[final]{changes} % Use for final version without revisions

\renewcommand*{\d}{\mathop{}\!\mathrm{d}}
\newcommand*{\e}{\mathop{}\!\mathrm{e}}

\newcommand{\integral}{\int\!\!}

\newcommand*{\trans}{{\mkern-1.5mu\mathsf{T}}}

\newcommand*{\qi}{\mathbf{i}}
\newcommand*{\qj}{\mathbf{j}}
\newcommand*{\qk}{\mathbf{k}}

\newcommand{\one}{\mathbbm{1}}
\newcommand{\lrn}{\overleftrightarrow{\nabla}}

\newcommand{\blrn}{\overleftrightarrow{\bm{\nabla}}}
\newcommand{\dl}{\varDelta}
\newcommand{\bd}{\bm{\varDelta}}
\newcommand{\bs}{\bm{\varSigma}}

\newcommand{\bohm}{\text{B}}
\newcommand{\conv}{\nabla}
\newcommand{\slnd}{\sigma}
\newcommand{\elem}{\text{elem.}}
\newcommand{\levy}{\text{LL}}

\newcommand{\bla}{\mathfrak{a}}
\newcommand{\blt}{\mathfrak{t}}
\newcommand{\ble}{\mathfrak{e}}

\definecolor{darkgreen}{RGB}{0,120,0}

\begin{document}

\title{
  Mechanical form factors and densities of non-relativistic fermions
}

\author{Adam Freese \orcidlink{0000-0002-0688-4121}\,}
\email{afreese@jlab.org}
\affiliation{Theory Center, Jefferson Lab, Newport News, Virginia 23606, USA}

\begin{abstract}
  The hadron physics community has been actively debating the interpretation
  of so-called mechanical properties of hadrons.
  Non-relativistic quantum-mechanical systems like
  the hydrogen atom have been appealed to in these debates as analogies.
  Since such appeals are likely to continue,
  it is important to have Galilei-covariant expressions
  for matrix elements of the energy-momentum tensor.
  In this work, I obtain Galilei-covariant breakdowns
  of such matrix elements into mechanical form factors,
  with a special focus on spin-half states.
  I additionally study the spatial densities associated with these form factors,
  using the pilot wave interpretation to guide their breakdown
  into contributions from internal structure and
  from quantum-mechanical effects such as wave packet dispersion.
  For completeness, I also obtain non-relativistic Breit frame densities.
\end{abstract}

\preprint{JLAB-THY-25-4317}

\maketitle

%%%%%%%%%%%%%%%%%%%%%%%%%%%%%%%%%%%%%%%%%%%%%%%%%%%%%%%%%%%%%%%%%%%%%%%%%%%%%%%%

% Intro
\section{Introduction}
\label{sec:intro}

In the past decade, the hadron physics community has seen a flourishing of discussion
about what are often called the mechanical properties of hadrons.
These are properties described by the energy-momentum tensor (EMT),
including the energy (or mass), angular momentum, and potentially internal forces.
Popular topics of discussion have included the decomposition of the proton's
mass~\cite{Ji:1995sv,Lorce:2017xzd,Metz:2020vxd,Ji:2021mtz,Lorce:2021xku}
and spin~\cite{Jaffe:1989jz,Ji:1996ek,Shore:1999be,Leader:2013jra},
as well as spatial distributions of
both~\cite{Polyakov:2002yz,Polyakov:2018zvc,Lorce:2018egm,Freese:2021czn,Panteleeva:2021iip,Freese:2021mzg,Freese:2022fat,Panteleeva:2022uii,Li:2024vgv,Won:2025dgc}.

Perhaps one of the most lively---and contentious---areas of discussion has been
the hadronic stress tensor.
The spacelike components of the EMT describe
momentum flux densities,
and in many continuum classical systems
have a clear interpretation as a stress tensor.
Since the seminal work of Maxim Polyakov~\cite{Polyakov:2002yz}---inspired largely
by analogy to liquid drops---many
researchers~\cite{Hudson:2017oul,Polyakov:2018zvc,Lorce:2018egm,Burkert:2018bqq,Kumericki:2019ddg,Cosyn:2019aio,Varma:2020crx,Metz:2021lqv,Freese:2021czn,Burkert:2021ith,Panteleeva:2021iip,Pefkou:2021fni,Freese:2021mzg,Lorce:2021xku,More:2021stk,Lorce:2022cle,Varma:2022kbv,Freese:2022fat,Panteleeva:2022uii,Duran:2022xag,Hackett:2023rif,Pefkou:2023okb,Hackett:2023nkr,Cao:2023ohj,GarciaMartin-Caro:2023toa,Hu:2024edc,Li:2024vgv,Dehghan:2025ncw,Goharipour:2025lep,Broniowski:2025ctl}
have utilized the traditionally classical continuum concepts
of stresses, pressures, tension and shear
in the interpretation of the hadronic EMT.
(For reviews, see Refs.~\cite{Polyakov:2018zvc,Burkert:2023wzr,Lorce:2025oot}.)
To be sure,
a few researchers~\cite{Ji:2021mfb,Ji:2022exr,Ji:2025gsq}
have questioned the concept of hadronic stresses;
but others~\cite{Burkert:2023wzr,Freese:2024rkr,Lorce:2025oot}
have argued to defend the use of the stress concept.

Many of these discussions use simpler systems than hadrons
as test cases for claims made in the literature.
After all, quantum chromodynamics is an infamously difficult theory.
Systems in quantum electrodynamics such as the
electron~\cite{Rodini:2020pis,Metz:2021lqv,Freese:2022jlu,Eides:2023uox}
and photon~\cite{Freese:2022ibw}
provide sandboxes to play with ideas often appealed to in discussions
of mechanical properties,
but in a better-understood theory.

Recently, the hydrogen atom has become a playground in which multiple
researchers~\cite{Ji:2022exr,Czarnecki:2023yqd,Eides:2024tzh,Freese:2024rkr,Fabbri:2025ffi}
have explored the calculation and meaning of the energy-momentum tensor.
As a non-relativistic quantum-mechanical system\footnote{
  Ref.~\cite{Fabbri:2025ffi} actually considers the hydrogen atom
  as a solution to the Dirac equation,
  and thus as a relativistic system.
},
it avoids difficulties related to relativistic effects in
spatial densities and composite operator renormalization.
With discussions about the meaning of the EMT far from over,
it is likely that appeals to non-relativistic quantum-mechanical systems
will continue.

If non-relativistic quantum mechanics is to become a playground
for studying mechanical properties,
it is important to have all the equipment and amenities in place.
Among these are expressions for matrix elements of the EMT in terms
of mechanical form factors\footnote{
  The form factors appearing in matrix elements of the EMT are more commonly
  called gravitational form factors,
  because the EMT is the source of gravitation in general relativity.
  However, I have found that the phrase ``gravitational form factors''
  confuses and misleads researchers outside the field,
  by giving the impression that gravitation is used to measure these
  form factors or that they are being used to study gravitational forces
  within hadrons.
  Since the purpose of these form factors is
  to characterize mechanical properties of hadrons---for the purpose
  of understanding the strong nuclear force, of course---it seems
  more apt to call them \emph{mechanical} form factors.
}.
Just as the guiding principle behind form factor breakdowns in
relativistic quantum field theory is Lorentz covariance,
the symmetries of non-relativistic physics---namely Galilei covariance---should
guide form factor breakdowns in non-relativistic quantum mechanics.
Such expressions seem to be absent in the present literature,
especially for systems with non-zero spin.
One of the purposes of this study is to provide a framework
for constructing such expressions,
and to provide form factor breakdowns for the EMT in
spin-zero and spin-half systems specifically.

Another helpful amenity to have on hand is a set of formulas
for relating the mechanical form factors to spatial densities
of mass, energy, momentum and stresses.
The other purpose of this work is to provide these density formulas.
Although the difficulties of relativistic effects are absent here,
there is still some ambiguity in separating the internal structure of composite
systems from effects of the barycentric wave packet (such as dispersion).
To this end, I use considerations from the pilot wave
interpretation~\cite{db:pilot,Bohm:1951xw,Bohm:1951xx,Bohm:2006und}
of quantum mechanics to guide this separation.
Nonetheless, for the benefit of readers who are skeptical of this interpretation,
I provide formulas for non-relativistic densities in the standard
Breit frame formalism as well.

This work is organized as follows.
In Sec.~\ref{sec:zero},
I provide a gentle introduction to the Galilei group,
using the energy-momentum tensor of spin-zero states as an example case.
In particular, I exploit the fact that the Galilei group in $3+1$ dimensions is
a subgroup of the $(4+1)$-dimensional Lorentz group.
Next, in Sec.~\ref{sec:spinor},
I dive more deeply into the relevant group theory,
to derive non-relativistic spinors that are covariant under the Galilei group.
The ultimate results of this deep dive are matrix elements of the spin-half EMT,
which are stated in Sec.~\ref{sec:mff}.
These are used to construct densities in Sec.~\ref{sec:pilot} with the guidance
of the pilot wave interpretation
(but contrasting density results in the Breit frame formalism are given in
Sec.~\ref{sec:breit} for completeness).
I conclude and provide an outlook in Sec.~\ref{sec:end}.

% Spin-zero states
\section{Spin-zero states}
\label{sec:zero}

In non-relativistic physics, we usually consider time $t$ and spatial coordinates
$\bm{x}$ separately, only uniting them into a single quantity---the four-vector
$x^\mu = (t;\bm{x})$---in relativistic physics\footnote{
  Here, and throughout the text, I use natural units with $c=1$.
  In the context of Galilei transformations,
  $c$ should be understood as merely a unit conversion factor.
}.
However, there is nothing to prevent us from formally considering $(t;\bm{x})$
as a unified object in non-relativistic physics---albeit with different
transformation laws than in special relativity.
Under an active boost of velocity $\bm{v}$ in particular:
\begin{align}
  \label{eqn:gal1}
  \begin{split}
    t'
    &=
    t
    \\
    \bm{x}'
    &=
    \bm{x} + \bm{v} t
    \,.
  \end{split}
\end{align}
Another familiar four-vector is the energy-momentum
four-vector $p^\mu = (E;\bm{p})$.
In relativistic physics, its components transform under the same
Lorentz transformation law as those of $x^\mu$.
What happens, however, in non-relativistic physics?
Under the same boost of velocity $\bm{v}$,
the energy and momentum of non-relativistic physics transform as:
\begin{align}
  \label{eqn:gal2}
  \begin{split}
    E'
    &=
    E
    +
    \bm{v}\cdot\bm{p}
    +
    \frac{1}{2}
    m \bm{v}^2
    \\
    \bm{p}'
    &=
    \bm{p} + \bm{v} m
    \,.
  \end{split}
\end{align}
At first glance, Eqs.~(\ref{eqn:gal1}) and (\ref{eqn:gal2})
seem to specify different transformation laws.
However, they can be unified---the hint for this
is the similar forms of the $\bm{x}$ and $\bm{p}$ transformation laws,
with $t$ and $m$ playing analogous roles.
If, rather than a four-vector,
we construct a five-vector:
\begin{align}
  \label{eqn:5vector}
  \begin{split}
    x^\mu
    &=
    (x^+; \bm{x}; x^-)
    =
    (t; \bm{x}; x^-)
    \\
    p^\mu
    &=
    (p^+; \bm{p}; p^-)
    =
    (m; \bm{p}; E)
    \,,
  \end{split}
\end{align}
with the two non-spatial components marked by an upper $+$ or $-$ index,
then these five-component objects transform identically under boosts:
\begin{align}
  \label{eqn:galileo}
  \begin{split}
    x^+
    &\mapsto
    x^+
    \\
    \bm{x}
    &\mapsto
    \bm{x}
    +
    \bm{v} x^+
    \\
    x^-
    &\mapsto
    x^-
    +
    \bm{v}\cdot\bm{x}
    +
    \frac{1}{2}
    x^+ \bm{v}^2
    \,.
  \end{split}
\end{align}
For the five-momentum in particular, the five components arise because the
energy and mass transform differently under boosts.
For the five-position on the other hand, the new component $x^-$ does not
have an apparent physical meaning,
and was only introduced so that $p^\mu$ and $x^\mu$
have the same transformation laws\footnote{
  A possible operational interpretation of the fifth coordinate
  has been suggested by Kapuscik~\cite{Kapuscik:1985nx},
  as a control parameter specifying the velocities of signals
  used to synchronize spatially-displaced clocks.
}.
In fact, five-vectors are widely used in studies of Galilei
symmetry~\cite{pinsky:gal,Kapuscik:1985nx,omote:gal,deMontigny:2003gdw,Santos:2004pq,deMontigny:2006,Niederle:2007xp,Abreu:2018ipp}.

Similarly to time and position,
the electric charge density $\rho(x)$ and current density $\bm{j}(x)$
are usually unified into a single
electromagnetic four-current in relativistic physics;
and similarly to time and position,
we can construct an electromagnetic five-current:
\begin{align}
  j^\mu(x)
  =
  ( \rho(x); \bm{j}(x); j^-(x) )
  \equiv
  ( j^+(x); \bm{j}(x); j^-(x) )
  \,.
\end{align}
Just as in the time-position case as well, $j^-(x)$ does not have a
clear physical meaning, and seems to be spurious.
The reason for identifying the charge density as the $+$ component
rather than the $-$ component is found in the transformation laws:
\begin{align}
  \begin{split}
    j^+(x)
    & \mapsto
    j^+(x')
    \\
    \bm{j}(x)
    & \mapsto
    \bm{j}(x')
    +
    \bm{v} j^+(x')
    \,.
  \end{split}
\end{align}
The general rule for five-currents is that the $+$ component is the density,
the $(1,2,3)$ components are the current density,
and the $-$ component is spurious and unphysical.

This brings us to the energy, momentum and mass currents.
In principle, the energy, mass, and all three components of the momentum
are locally conserved quantities which therefore
each have an associated five-current.
However, these five quantities are (unlike the charge)
not individually invariant under boosts,
being instead connected through the
Galilei transformation law (\ref{eqn:galileo}).
Thus, there should be a rank-two tensor $T^{\mu\nu}(x)$ for which
$T^{\mu +}(x)$ provides the mass current,
$T^{\mu a}(x)$ the momentum current (given $a\in\{1,2,3\}$)\footnote{
  Here, and throughout the text, I use $a,b,c$ instead of the usual $i,j,k$
  to signify spatial components of five-vectors.
},
and $T^{\mu -}(x)$ the energy current.
This tensor is in effect the non-relativistic analogue of the
relativistic energy-momentum tensor;
I will just call it the energy-momentum tensor through the remainder
of the text.

%%%%%%%%%%%%%%%%%%%%%%%%%%%%%%%%%%%%%%%%%%%%%%%%%%%%%%%%%%%%%%%%%%%%%%%%%%%%%%%%

\subsection{Mechanical form factors}
\label{sec:mech:zero}

Mechanical form factors are defined by breaking the matrix element
$\langle \bm{p}' | \hat{T}_q^{\mu\nu}(0) | \bm{p} \rangle$
down into the most general possible tensor structure that is compatible
with Galilei covariance
(and which is also invariant under parity and time reversal).
Here, $q$ specifies a particular constituent,
usually in the form of a particle species,
and $\hat{T}_q^{\mu\nu}$ signifies the contribution of this constituent to the EMT.
To build this tensor, we have as ingredients the five-vectors
$P^\mu = \frac{1}{2}\big(p^\mu+p'^\mu\big)$,
$\dl^\mu = p'^\mu - p^\mu$
and $n^\mu = (n^+;\bm{n};n^-) = (0;\bm{0},1)$,
as well as the Galilei metric tensor $g^{\mu\nu}$.
The vector $n^\mu$ is present in this list
because it is invariant under the Galilei group,
in effect representing that we could always add a constant to the
energy-like component of any five-vector without
altering the transformation law (\ref{eqn:galileo}).

The Galilei metric is defined so that Galilei transformations
are isometries of the metric---that is,
so that $g^{\mu\nu}$ is invariant under all Galilei transformations.
It must also be symmetric in its indices,
and by convention I choose $g^{ab}=-\delta^{ab}$ for $a,b\in\{1,2,3\}$
(to closely mimic similar components of the Minkowski metric).
These requirements impose the form:
\begin{align*}
  g^{\mu\nu}
  =
  \begin{bmatrix*}[r]
    0 &  0 &  0 &  0 & 1 \\
    0 & -1 &  0 &  0 & 0 \\
    0 &  0 & -1 &  0 & 0 \\
    0 &  0 &  0 & -1 & 0 \\
    1 &  0 &  0 &  0 & g^{--}
  \end{bmatrix*}
  \,
\end{align*}
where $g^{--}$ is an undetermined real number.
Choosing $g^{--}=0$,
in line with most authors~\cite{omote:gal,Santana:1998rd,deMontigny:2003gdw,Santos:2004pq,Abreu:2018ipp,deMontigny:2003gdw}\footnote{
  The early, pioneering work of Pinski~\cite{pinsky:gal}
  instead uses $g^{--}=1$.
},
conveniently makes the metric its own inverse:
\begin{align}
  \label{eqn:metric}
  g^{\mu\nu}
  =
  g_{\mu\nu}
  =
  \begin{bmatrix*}[r]
    0 &  0 &  0 &  0 & 1 \\
    0 & -1 &  0 &  0 & 0 \\
    0 &  0 & -1 &  0 & 0 \\
    0 &  0 &  0 & -1 & 0 \\
    1 &  0 &  0 &  0 & 0
  \end{bmatrix*}
  \,.
\end{align}
Additionally, $g^{\mu\nu}$ here is the metric tensor
of light front coordinates in five-dimensional spacetime.
This occurs because the Poincar\'{e} group has a Galilei subgroup
with one less dimension~\cite{pinsky:gal,Kogut:1969xa,Soper:1972xc,Burkardt:2002hr},
and is why I have used $+$ and $-$ superscripts for the null coordinates.

In light of the metric~(\ref{eqn:metric}),
scalar products between Galilei five-vectors are:
\begin{align}
  g_{\mu\nu}
  x^\mu y^\nu
  =
  x^+ y^-
  +
  x^- y^+
  -
  \bm{x}\cdot\bm{y}
  \,,
\end{align}
and the invariant associated with the mass-momentum-energy five-vector is:
\begin{align}
  \label{eqn:p2}
  g_{\mu\nu}
  p^\mu
  p^\nu
  =
  2 m E
  -
  \bm{p}^2
  =
  2 m E_0
  \,.
\end{align}
This will be useful later when constructing Galilei spinors.

In terms of the available tensors, the
matrix elements of the energy-momentum tensor takes the form:
\begin{align}
  \label{eqn:emt:zero}
  \langle \bm{p}' | \hat{T}_q^{\mu\nu}(0) | \bm{p} \rangle
  =
  \frac{P^\mu P^\nu}{m}
  A_q(\bd^2)
  +
  \frac{\dl^\mu \dl^\nu + g^{\mu\nu} \bd^2}{4m}
  D_q(\bd^2)
  +
  m
  g^{\mu\nu}
  \bar{c}_q(\bd^2)
  +
  P^\mu n^\nu
  \bar{e}_q(\bd^2)
  \,,
\end{align}
where $m$ is the mass of the system.
Strictly speaking, more tensors such as $n^\mu P^\nu$ and $n^\mu n^\nu$ could
be added, but these would contribute only to the unphysical components
$T^{-\nu}$ of the EMT, and can be safely discarded.
I have written this breakdown to closely mirror
the standard relativistic breakdown~\cite{Polyakov:2018zvc},
with the form factors $A_q(\bd^2)$ and $D_q(\bd^2)$ appearing
in various forms in works dating to the
1960s~\cite{Pagels:1966zza,Kobzarev:1962wt}
and $\bar{c}_q(\bd^2)$ first introduced by Ji in 1996~\cite{Ji:1996ek}.
The form factor $\bar{e}_q(\bd^2)$ is new to the non-relativistic case,
and is present because the energy density does not transform into
the mass and momentum densities under boosts----similarly to how
the rest energy in Eq.~(\ref{eqn:gal2}) does not transform into
the mass and momentum.
Another way to think about the $\bar{e}_q(\bd^2)$ form factor is that it
arises because Galilei symmetry does not include rotational symmetry
in the unphysical fifth dimension,
which makes it analogous to the ``spurious'' form factors appearing
in form factor breakdowns in light front
dynamics~\cite{Cao:2023ohj,Cao:2024rul}.

%%%%%%%%%%%%%%%%%%%%%%%%%%%%%%%%%%%%%%%%%%%%%%%%%%%%%%%%%%%%%%%%%%%%%%%%%%%%%%%%

\subsection{Sum rules}
\label{sec:sum}

There are several sum rules that can be imposed by the mechanical form factors
in Eq.~(\ref{eqn:emt:zero}).
First,
the densities of the energy-momentum tensor $T^{+\nu}$ should integrate
over all space to the five-momentum $P^\nu$.
Accordingly, once all constituents have been summed over, we should have:
\begin{align}
  \sum_q
  \langle \bm{p} | \hat{T}_q^{+\nu}(0) | \bm{p} \rangle
  =
  P^\nu
  \,.
\end{align}
This imposes the following sum rules on the form factors at $\bd=0$:
\begin{align}
  \begin{split}
    \sum_q A_q(0)
    &=
    1
    \\
    \sum_q
    \big(
    \bar{c}_q(0)
    +
    \bar{e}_q(0)
    \big)
    &=
    0
    \,.
  \end{split}
\end{align}
The first of these is analogous to the momentum sum rule for the relativistic form factors.
Additionally, the local continuity equation $\partial_\mu T^{\mu\nu}(\bm{x},t) = 0$ requires,
for non-zero momentum transfer:
\begin{align}
  \sum_q
  \dl_\mu
  \langle \bm{p}' | \hat{T}_q^{\mu\nu}(0) | \bm{p} \rangle
  \added{%
  =
  0%
  }
  \,,
\end{align}
and since \replaced{$\dl_\mu P^\mu=0$}{$\dl_\mu P^\nu = 0$}, this imposes the following sum rule:
\begin{align}
  \sum_q
  \bar{c}_q(\bd^2)
  =
  0
  \,,
\end{align}
which also holds for the relativistic $\bar{c}_q$ form factor~\cite{Ji:1996ek}.
Together with the previous sum rules, this entails:
\begin{align}
  \label{eqn:sum:ebar}
  \sum_q \bar{e}_q(0)
  =
  0
  \,.
\end{align}

%%%%%%%%%%%%%%%%%%%%%%%%%%%%%%%%%%%%%%%%%%%%%%%%%%%%%%%%%%%%%%%%%%%%%%%%%%%%%%%%

\subsection{Densities of the energy-momentum tensor}

With the form factor decomposition (\ref{eqn:emt:zero}),
we can construct its corresponding spatial densities.
The starting point is the expectation value of the local operator
$\hat{T}^{\mu\nu}(\bm{x})$ for physical states.
This expectation value can be written\footnote{
  See Refs.~\cite{Li:2022ldb,Freese:2022fat}
  for step-by-step derivations of similar expressions.
}:
\begin{align}
  \langle \Psi(t) | \hat{T}_q^{\mu\nu}(\bm{x}) | \Psi(t) \rangle
  &=
  \label{eqn:emt:zero:density}
  \integral \d^3 \bm{R}
  \integral \frac{\d^3\bd}{(2\pi)^3}
  \Psi^*(\bm{R},t)
  \langle \bm{p}' | \hat{T}_q^{\mu\nu}(0) | \bm{p} \rangle
  \Psi(\bm{R},t)
  \e^{-i\bd\cdot(\bm{R}-\bm{x})}
  \bigg|_{2\bm{P}\rightarrow-i\blrn}
  \,.
\end{align}
As far as specific components go:
the $\mu=+$ components constitute densities,
the $\mu\in\{1,2,3\}$ components constitute flux densities,
and the $\mu=-$ components have no physical meaning
(being akin to $j^-$).
The index $\nu$ labels which component of the five-momentum
this is a density or flux density of.
For the mass density,
mass flux density, and momentum density,
Eq.~(\ref{eqn:emt:zero:density}) factorizes into a single
internal density smeared out by different
wave-packet dependent smearing functions,
thus constituting ``simple'' densities in the nomenclature of
Ref.~\cite{Freese:2022fat}.
In particular, in terms of the probability density:
\begin{align}
  \mathscr{P}(\bm{R},t)
  &=
  \Psi^*(\bm{R},t)
  \Psi(\bm{R},t)
  \,,
\end{align}
the Bohmian velocity~\cite{Bohm:1951xw,Bohm:2006und}:
\begin{align}
  \bm{v}_{\bohm}(\bm{R},t)
  =
  -\frac{i}{2m}
  \frac{
    \Psi^*(\bm{R},t)
    \blrn
    \Psi(\bm{R},t)
  }{
    \Psi^*(\bm{R},t)
    \Psi(\bm{R},t)
  }
  \,,
\end{align}
and the contribution of the $q$th particle species to
the internal matter density:
\begin{align}
  \bla_q(\bm{b})
  &=
  \integral \frac{\d^3\bd}{(2\pi)^3}
  A_q(\bd^2)
  \e^{-i\bd\cdot\bm{b}}
  \,,
\end{align}
the expectation values in question can be written:
\begin{align}
  \begin{split}
    \langle \Psi(t) | \hat{T}_q^{++}(\bm{x}) | \Psi(t) \rangle
    &=
    m
    \integral \d^3 \bm{R} \,
    \mathscr{P}(\bm{R},t)
    \bla_q(\bm{x}-\bm{R})
    \\
    \langle \Psi(t) | \hat{T}_q^{a+}(\bm{x}) | \Psi(t) \rangle
    =
    \langle \Psi(t) | \hat{T}_q^{+a}(\bm{x}) | \Psi(t) \rangle
    &=
    m
    \integral \d^3 \bm{R} \,
    \mathscr{P}(\bm{R},t)
    v_{\bohm}^a(\bm{R},t)
    \,
    \bla_q(\bm{x}-\bm{R})
    \,.
  \end{split}
\end{align}
These expressions have a straightforward interpretation.
First, the average mass density is found by smearing the internal mass density
by the probability density $\mathscr{P}$.
Second, the momentum density is found by smearing instead by the probability current.
The second rule has a more transparent meaning in the pilot wave interpretation:
$\bm{v}_{\bohm}(\bm{R},t)$ is the composite system's actual velocity,
and a momentum distribution is obtained by multiplying this by the internal mass density.
The average momentum density is then obtained by smearing by the probability.

Next, the stress tensor constitutes a ``compound density''
in the nomenclature of Ref.~\cite{Freese:2022fat}:
\begin{align}
  \label{eqn:stress:zero}
  \begin{split}
    \langle \Psi(t) | \hat{T}_q^{ab}(\bm{x}) | \Psi(t) \rangle
    &=
    \integral \d^3 \bm{R} \,
    \left\{
      \left(
      -
      \Psi^*(\bm{R},t)
      \frac{\lrn^a \lrn^b}{4m}
      \Psi(\bm{R},t)
      \right)
      \bla_q(\bm{x} - \bm{R})
      +
      \mathscr{P}(\bm{R},t)
      \blt_q^{ab}(\bm{x}-\bm{R})
      \right\}
    \\
    \blt_q^{ab}(\bm{b})
    &=
    \integral \frac{\d^3\bd}{(2\pi)^3}
    \left\{
      \frac{\dl^a \dl^b - \bd^2 \delta^{ab}}{4m}
      D_q(\bd^2)
      -
      m \delta^{ab} \bar{c}_q(\bd^2)
      \right\}
    \e^{-i\bd\cdot\bm{b}}
    \,,
  \end{split}
\end{align}
where $\blt_q^{ab}(\bm{b})$ is the intrinsic stress tensor,
and the $\bla_q(\bm{b})$ term
constitutes dynamic stresses associated with wave packet motion and dispersion.
I previously have explored the meaning of this breakdown in the pilot wave
formulation~\cite{Freese:2024rkr}.

%%%%%%%%%%%%%%%%%%%%%%%%%%%%%%%%%%%%%%%%%%%%%%%%%%%%%%%%%%%%%%%%%%%%%%%%%%%%%%%%

\subsection{Energy density}
\label{sec:zero:energy}

The energy density is especially important,
being related to open questions and controversies
regarding mass generation and the mass decomposition of hadrons in
quantum chromodynamics
(see e.g., Refs.~\cite{Ji:1995sv,Lorce:2017xzd,Metz:2020vxd,Ji:2021mtz,Lorce:2021xku}).
Recalling that the energy is given by the ``minus'' component
of the Galilei five-momentum $p^-$,
the relevant matrix element is:
\begin{align}
  \label{eqn:zero:pm}
  \langle \bm{p}' | \hat{T}_q^{+-}(0) | \bm{p} \rangle
  =
  \left(
  E_0
  +
  \frac{\bm{P}^2}{2m}
  +
  \frac{ \bd^2 }{8m}
  \right)
  A_q(\bd^2)
  +
  \frac{ \bd^2 }{4m}
  D_q(\bd^2)
  +
  m
  \big(
  \bar{c}_q(\bd^2)
  +
  \bar{e}_q(\bd^2)
  \big)
  \,,
\end{align}
where $E_0 \approx m$ is the rest energy of the system\footnote{
  Recall that in non-relativistic mechanics,
  mass is strictly additive:
  for a system of $N$ particles,
  $m = m_1 + m_2 + \ldots + m_N$.
  However, even if we ascribe a rest energy $m_n$
  to each of these particles when they are free,
  the energy of the system in the center-of-mass frame
  still contains binding and kinetic energy contributions,
  and thus $E_0 \neq m$.
  Of course since the motion of all constituents is assumed
  (by the very applicability of non-relativistic physics)
  to be much slower than the speed of light,
  these deviations of $E_0$ from $m$ are small,
  so $E_0 \approx m$.
}.
The corresponding energy density will naturally be a compound density,
with one contribution from the energy contained in barycentric motion
and wave packet dispersion,
and the other from the internal density.

It may be tempting to identify $\frac{\bm{P}^2}{2m} A_q(\bd^2)$ as containing
all of the barycentric energy,
and the remaining terms as the internal energy---and
this is in effect what the Breit frame formalism
does~\cite{Polyakov:2018zvc}.
Perhaps in vanilla quantum mechanics,
the separation of the energy density into barycentric and internal contributions
is to some extent arbitrary, and thus a matter of definition.
However, in the realist pilot wave interpretation,
there is a clean and unambiguous way of separating the barycentric
and internal contributions to the energy density.
While the truth of the pilot wave formulation is not evident---and
empirically, cannot be, since it makes the same predictions
as vanilla quantum theory---there is no harm in hedging our bets on the matter,
and adopting a convention that \emph{might} turn out to be the true
breakdown under a realist interpretation of quantum mechanics.

In the pilot wave interpretation,
the system as a whole has a kinetic energy $K$ and a quantum potential energy $Q$,
given respectively by~\cite{Bohm:1951xw,Bohm:2006und}:
\begin{align}
  \begin{split}
    K(\bm{R},t)
    &=
    \frac{(\bm{\nabla} \mathscr{S}(\bm{R},t))^2}{2m}
    \\
    Q(\bm{R},t)
    &=
    -
    \frac{1}{2m}
    \frac{\bm{\nabla}^2\mathscr{R}(\bm{R},t)}{\mathscr{R}(\bm{R},t)}
    \,,
  \end{split}
\end{align}
where $\mathscr{R}(\bm{R},t)$ and $\mathscr{S}(\bm{R},t)$ are
real-valued functions
defined through the polar decomposition of the wave function:
\begin{align}
  \Psi(\bm{R},t)
  =
  \mathscr{R}(\bm{R},t)
  \e^{i \mathscr{S}(\bm{R},t)}
  \,.
\end{align}
The sum of the kinetic and quantum potential energy,
weighted by the probability density, can be written:
\begin{align}
  \mathscr{E}_{\mathrm{bary}}(\bm{R},t)
  \equiv
  \mathscr{P}(\bm{R},t)
  \big( K(\bm{R},t) + Q(\bm{R},t) \big)
  =
  -\frac{1}{8m}
  \Psi^*(\bm{R},t)
  \blrn^2
  \Psi(\bm{R},t)
  -
  \frac{1}{8m}
  \bm{\nabla}^2
  \big[
    \Psi^*(\bm{R},t)
    \Psi(\bm{R},t)
    \big]
  \,,
\end{align}
where I have identified this energy as ``bary'' for barycentric.
The two-sided derivative in this expression
translates to $-4\bm{P}^2$ in momentum space,
and the total Laplacian translates to $-\bd^2$.
Thus the barycentric energy involves not just the $\bm{P}^2$
term in the form factor breakdown (\ref{eqn:zero:pm}),
but also part of the $\bd^2$ terms.
In fact, the resulting energy density breakdown is:
\begin{align}
  \langle \Psi(t) | \hat{T}^{+-}(\bm{x}) | \Psi(t) \rangle
  &=
  \integral \d^3 \bm{R} \,
  \Big\{
    \mathscr{E}_{\mathrm{bary}}(\bm{R},t)
    \bla_q(\bm{x} - \bm{R})
    +
    \mathscr{P}(\bm{R},t)
    \ble_q(\bm{x}-\bm{R})
    \Big\}
  \\
  \label{eqn:energy:zero}
  \ble_q(\bm{b})
  &=
  \integral \frac{\d^3\bd}{(2\pi)^3}
  \left\{
    E_0
    A_q(\bd^2)
    +
    m
    \big(
    \bar{c}_q(\bd^2)
    +
    \bar{e}_q(\bd^2)
    \big)
    +
    \frac{ \bd^2 }{4m}
    D_q(\bd^2)
    \right\}
  \e^{-i\bd\cdot\bm{b}}
  \,,
\end{align}
where $\ble_q(\bm{b})$ is the contribution of the
$q$th particle species to the internal energy density.

A few remarks about this result are in order.
Existing formulas
for the internal energy density of \added{spin-zero} composite systems~\cite{Polyakov:2018zvc,Freese:2022fat}
\deleted{(see Refs.~\cite{} for examples)}
\deleted{all} contain a term where $\bd^2$ multiplies $A_q(\bd^2)$.
This means, even for an elementary particle with no $D$-term---for
which $A(\bd^2)=1$ and $\bar{c}(\bd^2)=0$---there is a non-trivial
internal energy density.
This is peculiar, given that the particle in question is ostensibly pointlike.
By contrast, the result (\ref{eqn:energy:zero})
gives a delta function times $E_0$ for such an elementary particle.
While this by no means proves the correctness of
Eq.~(\ref{eqn:energy:zero}) over standard formulas,
it does make the result more intuitive.

It's also worth noting that the integral of
Eq.~(\ref{eqn:energy:zero}) over all space gives:
\begin{align}
  \integral \d^3 \bm{b} \,
  \ble_q(\bm{b})
  =
  E_0
  A_q(0)
  +
  m
  \big(
  \bar{c}_q(0)
  +
  \bar{e}_q(0)
  \big)
  \,,
\end{align}
making this the contribution of the $q$th constituent to the rest energy.
This is in effect the non-relativistic analogue of
Lorc\'{e}'s rest energy decomposition~\cite{Lorce:2017xzd},
which in the relativistic case gives the constituent contribution as
$m\big(A_q(0) + \bar{c}_q(0)\big)$.

%%%%%%%%%%%%%%%%%%%%%%%%%%%%%%%%%%%%%%%%%%%%%%%%%%%%%%%%%%%%%%%%%%%%%%%%%%%%%%%%

\subsection{Covariant breakdown}

Before moving on to spin-half states,
it is helpful to unify the densities I have written above in an apparently
covariant form.
To this end, it is helpful to define the following Galilei boost
matrices~\cite{deMontigny:2003gdw}:
\begin{align}
  \label{eqn:boost}
  \varLambda^{\mu}_{\phantom{\mu}\nu}(\bm{v})
  =
  \begin{bmatrix}
    1 & 0 & 0 & 0 & 0 \\
    v_x & 1 & 0 & 0 & 0 \\
    v_y & 0 & 1 & 0 & 0 \\
    v_z & 0 & 0 & 1 & 0 \\
    \frac{1}{2}\bm{v}^2 & v_x & v_y & v_z & 1
  \end{bmatrix}
  \,.
\end{align}
Using these,
the expectation value of the energy-momentum tensor can be written:
\begin{align}
  \label{eqn:full:zero}
  \langle \Psi(t) | \hat{T}_q^{\mu\nu}(\bm{x}) | \Psi(t) \rangle
  &=
  \integral \d^3 \bm{R} \,
  \varLambda^{\mu}_{\phantom{\mu}\alpha}\big(\bm{v}_{\bohm}(\bm{R},t)\big)
  \varLambda^{\nu}_{\phantom{\nu}\beta} \big(\bm{v}_{\bohm}(\bm{R},t)\big)
  \left\{
    \mathscr{P}(\bm{R},t)
    \blt_q^{\alpha\beta}(\bm{x}-\bm{R})
    +
    T_Q^{\alpha\beta}(\bm{R},t)
    \bla_q(\bm{x}-\bm{R})
    \right\}
  \,,
\end{align}
where $\blt_q^{\alpha\beta}(\bm{b})$ is the internal energy-momentum tensor,
and
$T_Q^{\alpha\beta}(\bm{R},t)$ is the quantum energy-momentum tensor.
The internal energy-momentum tensor has components including:
\begin{align}
  \blt_q^{++}(\bm{b})
  =
  \bla_q(\bm{b})
  \qquad
  \qquad
  \blt_q^{+-}(\bm{b})
  =
  \ble_q(\bm{b})
  \,,
\end{align}
as well as those given in Eq.~(\ref{eqn:stress:zero}).
These can be interpreted as mass, energy and momentum currents
inside the system, owing to its composite structure,
in its rest frame.

The quantum energy-momentum tensor, on the other hand,
has the following non-zero components:
\begin{align}
  \label{eqn:emt:quantum}
  \begin{split}
    T_{Q}^{ab}(\bm{R},t)
    &=
    \frac{1}{2m}
    \Big(
    \big(\nabla_a\mathscr{R}(\bm{R},t)\big)
    \big(\nabla_b\mathscr{R}(\bm{R},t)\big)
    -
    \mathscr{R}(\bm{R},t)
    \big(\nabla_a \nabla_b \mathscr{R}(\bm{R},t)\big)
    \Big)
    \\
    T_{Q}^{+-}(\bm{R},t)
    &=
    -\frac{
      \mathscr{R}(\bm{R},t)
      \bm{\nabla}^2\mathscr{R}(\bm{R},t)
    }{2m}
    \\
    T_{Q}^{a-}(\bm{R},t)
    &=
    \frac{1}{2m}
    \left(
    \frac{\partial\mathscr{R}(\bm{R},t)}{\partial t}
    \big(\nabla_a\mathscr{R}(\bm{R},t)\big)
    -
    \mathscr{R}(\bm{R},t)
    \nabla_a\left[
      \frac{\partial\mathscr{R}(\bm{R},t)}{\partial t}
      \right]
    \right)
  \end{split}
\end{align}
the first line of which was first given by Takabayasi in Eq.~(A2) of Ref.~\cite{Takabayasi:1952}.
These components effectively describe stresses and energies
felt by the composite particle due to its guidance by the barycentric wave function
in the particle's rest frame.

Finally, the contributions of both the intrinsic and quantum EMT in
Eq.~(\ref{eqn:full:zero})
are boosted by the Bohmian velocity
$\bm{v}_{\bohm}(\bm{R},t)$,
which effectively incorporates contributions of barycentric motion
to the energy and momentum densities and fluxes.
This is then in turn smeared out by the probability $\mathscr{P}(\bm{R},t)$
of the barycentric position actually being $\bm{R}$.

% Galilean spinors
\section{Galilei spinors}
\label{sec:spinor}

With spin-zero states under our belt,
the time is ripe to move on to spin-half states.
To construct a spin-half analogue of Eq.~(\ref{eqn:emt:zero}),
we need to write the matrix element
$\langle \bm{p}', s' | \hat{T}^{\mu\nu}(0) | \bm{p}, s \rangle$
in the most general possible manner compatible with Galilei covariance
(as well as parity and time reversal),
using all the five-vectors and Galilei tensors at our disposal.
Since it is a spin-half state,
the available structures include Clifford algebra matrices,
and everything should be sandwiched between
Galilei-covariant spinors $u(\bm{p},s)$.

There is one difficulty which will take some time to address,
and which is the focus of the current section:
we need the appropriate Galilei-covariant spinors to do this.
The popular two-component Pauli spinors are inadequate for this purpose,
because they do not transform under a matrix representation of
the Galilei group~\cite{Huegele:2011}.
There is a standard set of four-component
spinors discovered by L\'evy-Leblond~\cite{Levy-Leblond:1967eic}
that are covariant under the part of the Galilei group connected to identity.
However, they do not have the expected transformation properties
under parity reversal,
which limits their applicability to the construction of
form factor breakdowns.
(See Appendix~\ref{sec:levyleblond} for details.)
This necessitates the derivation of new Galilei-covariant spinors.

In this section,
I will construct an eight-component spinor wave equation that is
covariant under the full Galilei group,
and give explicit solutions\footnote{
  It should be noted that an eight-component spinor wave equation
  was derived previously by
  Kobayashi, de Montigny and Khanna~\cite{Kobayashi:2007jn},
  by a Clifford algebra method in $5+1$ dimensions,
  but I will obtain simpler spinors directly in
  $4+1$ dimensions using representation theory of
  the $(4+1)$-dimensional Lorentz group.
}.
The construction will be by necessity rather mathematically involved.
A reader only interested in results
can skip to Sec.~\ref{sec:explicit},
where I write the covariant form factor breakdown and evaluate its components
in terms of standard (but non-covariant) two-component spinors.

%%%%%%%%%%%%%%%%%%%%%%%%%%%%%%%%%%%%%%%%%%%%%%%%%%%%%%%%%%%%%%%%%%%%%%%%%%%%%%%%

\subsection{Representation theory for the $(4+1)$-dimensional Lorentz group}

To derive the required spinors,
I will step back and consider representation theory of the
$(4+1)$-dimensional Lorentz group.
This group contains the connected part of the
$(3+1)$-dimensional Galilei group as a subgroup,
so spinors in $(4+1)$-dimensional Minkowski spacetime will be
covariant under the Galilei group after restricting $p^+=m$.
A subtlety that arises along the way is that Galilean parity reversal
is not present in the $(4+1)$-dimensional Lorentz group,
so the Galilei subgroup contained therein will need to be extended,
but we'll cross that bridge when we come to it.

The Lorentz group in $(4+1)$ dimensions is the special pseudo-orthogonal group
$\mathrm{SO}(4,1,\mathbb{R})$
of matrices that preserve the metric
$g = \mathrm{diag}(1,-1,-1,-1,-1)$, i.e.,
\begin{align}
  M g M^\trans
  =
  g
  \,,
\end{align}
and have a determinant $\mathrm{det}(M) = 1$.
The subgroup that is path-connected to identity
can be written in terms of 10 generators
$\lambda^{\mu\nu} = -i J^{\mu\nu}$
(4 boosts and 6 rotations),
with $\lambda_{\mu\nu}$ being antisymmetric in its indices.
Given a set of 10 real-valued parameters $\omega_{\mu\nu}$
(also antisymmetric):
\begin{align}
  M
  =
  \exp\left\{
    \frac{1}{2} \omega_{\mu\nu} \lambda^{\mu\nu}
    \right\}
  =
  \exp\left\{
    -\frac{i}{2} \omega_{\mu\nu} J^{\mu\nu}
    \right\}
  \,,
\end{align}
where the factor
$\frac{1}{2}$ is to counteract double-counting.
It is common for mathematicians to use exponentiation without a factor $i$,
and for physicists to use the factor $i$,
making the conventions for the generators differ by an imaginary unit.
In this section I will use the mathematicians' convention.

The algebra $\mathfrak{so}(4,1,\mathbb{R})$ of the generators
is specified by the commutation rules:
\begin{align}
  \label{eqn:algebra}
  [\lambda_{\mu\nu}, \lambda_{\rho\sigma}]
  =
  g_{\mu\sigma}
  \lambda_{\nu\rho}
  +
  g_{\nu\rho}
  \lambda_{\mu\sigma}
  -
  g_{\mu\rho}
  \lambda_{\nu\sigma}
  -
  g_{\nu\sigma}
  \lambda_{\mu\rho}
  \,,
\end{align}
or alternatively:
\begin{align}
  [J_{\mu\nu}, J_{\rho\sigma}]
  =
  i
  \Big(
  g_{\mu\sigma}
  J_{\nu\rho}
  +
  g_{\nu\rho}
  J_{\mu\sigma}
  -
  g_{\mu\rho}
  J_{\nu\sigma}
  -
  g_{\nu\sigma}
  J_{\mu\rho}
  \Big)
  \,,
\end{align}
the latter being the Lorentz algebra commutation relation
familiar to most physicists~\cite{Itzykson:1980rh,Weinberg:1995mt}.
The components $J_{0i}$ (or $\lambda_{0i}$)
can be identified as boost generators in particular,
and $J_{ij}$ (or $\lambda_{ij}$)
as generators of rotations in the $(x_i,x_j)$ plane.

%%%%%%%%%%%%%%%%%%%%%%%%%%%%%%%%%%%%%%%%

\subsubsection{Double cover of the five-dimensional Lorentz group}
\label{sec:spinor:cover}

Spinors are objects that transform under the double-cover of the Lorentz group.
In four-dimensional spacetime, the Lorentz group
$\mathrm{SO}(3,1,\mathbb{R})$
has $\mathrm{SL}(2,\mathbb{C})$ as its double cover.
Left-handed and right-handed spinors then transform under two inequivalent
fundamental representations of
$\mathrm{SL}(2,\mathbb{C})$.

In analogy, the five-dimensional Lorentz group also has a double cover:
the pseudo-unitary group $\mathrm{U}(1,1,\mathbb{H})$,
of $2\times2$ quaternion-valued matrices that preserve the metric
$\eta = \mathrm{diag}(1,-1)$~\cite{Takahashi:1963,Strom:1970,gilmore2006lie}:
\begin{align}
  \label{eqn:U11H}
  M \eta M^\dagger
  =
  \eta
  \,.
\end{align}
Alternatively, one can avoid the use of quaternions
by using the pseudo-unitary symplectic group $\mathrm{USp}(2,2,\mathbb{C})$
of complex $4\times4$ matrices that satisfy both~\cite{enayati2024sitter,Pejhan:2023rbt}:
\begin{align}
  \begin{array}{ll}
  M
  H
  M^\dagger
  =
  H
  \qquad \qquad % padding
    &
  \qquad \qquad % padding
  H
  =
  \mathrm{diag}(1,1,-1,-1)
  \\
  M \varOmega M^\trans
  =
  \varOmega
  \qquad \qquad % padding
    &
  \qquad \qquad % padding
  \varOmega
  =
  \begin{bmatrix}
    i\sigma_2 & 0 \\
    0 & i\sigma_2
  \end{bmatrix}
  \end{array}
\end{align}
which is isomorphic to $\mathrm{U}(1,1,\mathbb{H})$~\cite{gilmore2006lie,enayati2024sitter,Pejhan:2023rbt}.
The isomorphism maps the unit quaternions $(\qi,\qj,\qk)$ to the Pauli matrices:
\begin{align}
  \label{eqn:map}
  \qi
  \equiv
  -i\sigma_1
  \qquad
  \qj
  \equiv
  -i\sigma_2
  \qquad
  \qk
  \equiv
  -i\sigma_3
  \,.
\end{align}
Ultimately, the complex matrix representation will be necessary to construct
explicit spinors,
but quaternion algebra allows faster rote calculations,
and the map (\ref{eqn:map}) allows
the result of any quaternion calculation to be promptly
translated into the complex matrix representation.

The generators of $\mathrm{U}(1,1,\mathbb{H})$ can be written:
\begin{align}
  \label{eqn:gen:q}
  \begin{split}
    %%%%%%%%%%%%%%%%%%
    &
    \lambda_{01}
    =
    \frac{1}{2}
    \begin{bmatrix*}[r]
      0 & -\qi \\
      \qi & 0
    \end{bmatrix*}
    \qquad
    \lambda_{02}
    =
    \frac{1}{2}
    \begin{bmatrix*}[r]
      0 & -\qj \\
      \qj & 0
    \end{bmatrix*}
    \qquad
    \lambda_{03}
    =
    \frac{1}{2}
    \begin{bmatrix*}[r]
      0 & -\qk \\
      \qk & 0
    \end{bmatrix*}
    \qquad
    \lambda_{04}
    =
    \frac{1}{2}
    \begin{bmatrix}
      0 & 1 \\
      1 & 0
    \end{bmatrix}
    %%%%%%%%%%%%%%%%%%
    \\
    &
    \lambda_{14}
    =
    \frac{1}{2}
    \begin{bmatrix*}[r]
      \qi & 0 \\
      0 & -\qi
    \end{bmatrix*}
    \qquad
    \lambda_{24}
    =
    \frac{1}{2}
    \begin{bmatrix*}[r]
      \qj & 0 \\
      0 & -\qj
    \end{bmatrix*}
    \qquad
    \lambda_{34}
    =
    \frac{1}{2}
    \begin{bmatrix*}[r]
      \qk & 0 \\
      0 & -\qk
    \end{bmatrix*}
    %%%%%%%%%%%%%%%%%%
    \\
    &
    \lambda_{23}
    =
    \frac{1}{2}
    \begin{bmatrix*}[r]
      \qi & 0 \\
      0 & \phantom{-}\qi % alignment with matrix in previous row
    \end{bmatrix*}
    \qquad
    \lambda_{31}
    =
    \frac{1}{2}
    \begin{bmatrix*}[r]
      \qj & 0 \\
      0 & \phantom{-}\qj % alignment with matrix in previous row
    \end{bmatrix*}
    \qquad
    \lambda_{12}
    =
    \frac{1}{2}
    \begin{bmatrix*}[r]
      \qk & 0 \\
      0 & \phantom{-}\qk % alignment with matrix in previous row
    \end{bmatrix*}
    \,,
  \end{split}
\end{align}
and obey the commutation relations (\ref{eqn:algebra}).
From these generators, one can readily confirm through exponentiation
that a $2\pi$ rotation in any spatial plane
produces a factor $-1$, as expected for fermion representations.

Unlike the four-dimensional case,
there is only one fundamental representation of the covering group
up to unitary equivalence.
In the four-dimensional case,
right- and left-handed spinors transform under
$\psi_R \mapsto M\psi_R$
and
$\psi_L \mapsto (M^\dagger)^{-1}\psi_L$
respectively,
and there is no unitary $U$ for which $(M^\dagger)^{-1} = UMU^\dagger$
for every $M\in\mathrm{SL}(2,\mathbb{C})$.
By contrast, for any $M \in \mathrm{U}(1,1,\mathbb{H})$,
$\eta = \mathrm{diag}(1,-1)$ is a unitary matrix for which:
\begin{align}
  \eta
  M
  \eta
  =
  (M^\dagger)^{-1}
  \,,
\end{align}
as in fact follows from the definition of the group~(\ref{eqn:U11H}).
This means, for instance,
if a spinor transforms as $\psi \mapsto M\Psi$,
then $\eta\psi \mapsto (M^\dagger)^{-1}\eta\psi$.
On the other hand, as we shall see later,
the $M$ and $(M^\dagger)^{-1}$ representations are no longer equivalent
when the group is extended to include Galilean parity reversal.

%%%%%%%%%%%%%%%%%%%%%%%%%%%%%%%%%%%%%%%%

\subsubsection{Matrix representation of five-vectors}
\label{sec:spinor:vector}

A helpful trick often used in four-dimensional relativistic
physics is to write
four-vectors as $2\times 2$ complex-valued matrices through the construction
$\hat{X} = X^\mu \sigma_\mu$,
where $\sigma_\mu = (\one;\sigma_1,\sigma_2,\sigma_3)$
are the Infeld-van der Waerden symbols~\cite{Infeld:1933zz,Cosyn:2025gmp}.
This is possible because four-vectors are rank-two spinors,
with indices that transform under the left-handed and right-handed defining
representations of $\mathrm{SL}(2,\mathbb{C})$.
(See Refs.~\cite{waerden:1929,*groningen2017spinoranalysis,Penrose_Rindler_1984,Haag:1992hx,Srednicki:2007qs,Dreiner:2008tw} for further details.)
A similar trick can be employed for five-vectors---which
are similarly rank-two spinors with respect to
the defining representation of $\mathrm{U}(1,1,\mathbb{H})$,
and accordingly can be represented using
$2\times 2$ quaternion-valued matrices.
The appropriate construction is~\cite{Strom:1970}:
\begin{align}
  \label{eqn:vm}
  \begin{split}
    &
    \hat{X}
    =
    X^\mu \tau_\mu
    =
    \begin{bmatrix}
      X^0
      &
      X^4 - X^1 \qi - X^2 \qj - X^3 \qk
      \\
      X^4 + X^1 \qi + X^2 \qj + X^3 \qk
      &
      X^0
    \end{bmatrix}
    \\
    &
    \tau_0
    =
    \begin{bmatrix}
      1 & 0 \\
      0 & 1
    \end{bmatrix}
    \qquad
    %
    %&
    \tau_1
    =
    \begin{bmatrix*}[r]
      0 & -\qi \\
      \qi & 0
    \end{bmatrix*}
    \qquad
    \tau_2
    =
    \begin{bmatrix*}[r]
      0 & -\qj \\
      \qj & 0
    \end{bmatrix*}
    \qquad
    \tau_3
    =
    \begin{bmatrix*}[r]
      0 & -\qk \\
      \qk & 0
    \end{bmatrix*}
    \qquad
    \tau_4
    =
    \begin{bmatrix}
      0 & 1 \\
      1 & 0
    \end{bmatrix}
    \,.
  \end{split}
\end{align}
These $\tau_\mu$ matrices are in effect a five-dimensional analogue
of the Infeld-van der Waerden symbols $\sigma_\mu$.
Under a five-dimensional Lorentz transformation,
the five-vector transforms as:
\begin{align}
  \label{eqn:vm:trans}
  \hat{X}
  \mapsto
  \hat{X}'
  =
  M \hat{X} M^\dagger
  \,.
\end{align}
One can verify by explicit calculation that this gives
the expected transformation laws.

%%%%%%%%%%%%%%%%%%%%%%%%%%%%%%%%%%%%%%%%

\subsubsection{The Galilei subgroup}

The matrix construction (\ref{eqn:vm}) for five-vectors
is especially helpful for identifying the Galilei subgroup.
This is the subgroup generated by rotations in the $xy$, $yz$ and $zx$
planes, along with boosts following the Galilei transformation rule
(\ref{eqn:galileo}), where
$x^\pm = \frac{1}{\sqrt{2}}\big(x^0 \pm x^4\big)$.
The matrix representation of five-vectors (\ref{eqn:vm})
can be written in terms of light front coordinates if:
\begin{align}
  \tau^-
  =
  \tau_+
  =
  \frac{1}{\sqrt{2}}
  \begin{bmatrix}
    1 & 1 \\
    1 & 1
  \end{bmatrix}
  \qquad
  \qquad
  \tau^+
  =
  \tau_-
  =
  \frac{1}{\sqrt{2}}
  \begin{bmatrix*}[r]
    1 & -1 \\
    -1 & 1
  \end{bmatrix*}
  \,.
\end{align}
A helpful rule of thumb for identifying Galilei transformations
is that they will not transform $\tau_-$ into other tau matrices,
and they will not transform other tau matrices into $\tau_+$---ensuring
that Galilean time is invariant and Galilean energy
does not mix into momentum or mass.
One can show through explicit calculation
that the Galilei subgroup can be generated by:
\begin{align}
  \begin{split}
    &
    \lambda_{23} \,,
    \qquad
    \lambda_{31} \,,
    \qquad
    \lambda_{12} \,,
    \\
    &
    \beta_{a}
    =
    \frac{1}{\sqrt{2}}
    \big(
    \lambda_{0a} + \lambda_{4a}
    \big)
    \qquad
    a \in \{1,2,3\}
    \,,
  \end{split}
\end{align}
with the $\lambda_{ab}$ constituting rotations and the $\beta_a$ constituting boosts.
The Galilei boosts are similar in form to the light front transverse boosts
in Refs.~\cite{Kogut:1969xa,Soper:1972xc,Burkardt:2002hr},
the three-dimensional $xyz$-space being the ``transverse plane''
in this case.
Using the map (\ref{eqn:map}),
the complex matrix representation for the Galilei boost generators is:
\begin{align}
  \label{eqn:boost:2x2}
  \beta_a
  =
  \frac{i}{2\sqrt{2}}
  \begin{bmatrix}
    \sigma_a & \sigma_a \\
    -\sigma_a & -\sigma_a
  \end{bmatrix}
  \,,
\end{align}
which will be indispensable for constructing the Galilei spinors.

%%%%%%%%%%%%%%%%%%%%%%%%%%%%%%%%%%%%%%%%%%%%%%%%%%%%%%%%%%%%%%%%%%%%%%%%%%%%%%%%

\subsection{The five-dimensional Dirac equation}
\label{sec:dirac}

The Dirac equation in five dimensions involves eight-component spinors,
since there are
are positive- and negative-energy solutions
(i.e., particles and anti-particles)
that can have simultaneous eigenvalues of $\pm\frac{1}{2}$
for both $J_{12}$ and $J_{34}$.
While this may seem counterintuitive from familiarity with
three-dimensional space,
rotations in the $xy$ and $zw$ planes commute in four spatial dimensions.
The positive- and negative-energy solutions should thus each
have four independent components, giving eight total.

The five-dimensional Dirac equation
(with $p^\mu p_\mu = 2mE_0$ as invariant) is:
\begin{align}
  \label{eqn:dirac}
  p_\mu
  \gamma^\mu
  u(p)
  \equiv
  p^\mu
  \begin{bmatrix}
    0 & \tau_\mu \\
    \eta \tau_\mu \eta & 0
  \end{bmatrix}
  \begin{bmatrix}
    \chi_1(p) \\ \eta\chi_2(p)
  \end{bmatrix}
  =
  \sqrt{2 m E_0}
  \begin{bmatrix}
    \chi_1(p) \\ \eta\chi_2(p)
  \end{bmatrix}
  \,,
\end{align}
which effectively defines the $8\times 8$ gamma matrices,
which are explicitly given in Appendix~\ref{sec:gamma}.
Each of $\chi_{1,2}(p)$ is a four-component column matrix in the complex defining
representation of the covering group.
In line with the discussion in Sec.~\ref{sec:spinor:cover},
the top four components of $u(p)$ transform as
$\chi_1(p) \mapsto M \chi_1(p)$,
while the bottom four components transform as
$\eta\chi_2(p) \mapsto (M^\dagger)^{-1} \eta \chi_2(p)$.
The full eight-component spinor thus transforms as:
\begin{align}
  \label{eqn:spinor:trans}
  u(p)
  \mapsto
  \begin{bmatrix}
    M & 0 \\
    0 & (M^\dagger)^{-1}
  \end{bmatrix}
  u(p)
  \,.
\end{align}
Given the transformation law (\ref{eqn:vm:trans}),
it is straightforward to verify that the Dirac equation (\ref{eqn:dirac})
is covariant under the full $(4+1)$-dimensional Lorentz group---and thus,
accordingly, under its Galilei subgroup.

One apparent disanalogy between the usual Dirac spinors and this construction
is that Dirac spinors in $(3+1)$ dimensions are built from inequivalent defining
representations of $\mathrm{SL}(2,\mathbb{C})$,
whereas here there is only one defining representation of $\mathrm{U}(1,1,\mathbb{H})$.
However, $\mathrm{U}(1,1,\mathbb{H})$ does not include discrete Galilei transformations
such as parity reversal.
It turns out that the four-component spinors $\chi_1(p)$ and $\eta\chi_2(p)$ transform
inequivalently under an extended group that includes Galilean parity reversal.
I will address this presently.

%%%%%%%%%%%%%%%%%%%%%%%%%%%%%%%%%%%%%%%%

\subsubsection{Parity reversal}
\label{sec:parity}

Let us next consider how the new spinors transform under
\emph{Galilean} parity inversion.
This will be vital when restricting ourselves to the Galilei subgroup,
to ensure the physical solutions we construct transform correctly under parity.

\emph{Galilean} parity is distinct from Lorentzian parity in this context,
and should transform only the $(1,2,3)$ components of
five-vectors, leaving the $(+,-)$ components
(and thus, the $(0,4)$ components) alone:
\begin{align}
  \begin{split}
    P \tau^{1,2,3} (P^\dagger)^{-1}
    &=
    - \tau^{1,2,3}
    \\
    P \tau^{0,4} (P^\dagger)^{-1}
    &=
    + \tau^{0,4}
    \,.
  \end{split}
\end{align}
A matrix that satisfies these is $M = (M^\dagger)^{-1} = \tau_4$.
Given Eq.~(\ref{eqn:spinor:trans}),
the eight-component spinors transform under Galilean parity reversal as:
\begin{align}
  \label{eqn:parity}
  P
  =
  \begin{bmatrix}
    \tau_4 & 0 \\
    0 & \tau_4
  \end{bmatrix}
  =
  \begin{bmatrix}
    0 & \one_{2\times2} & 0 & 0 \\
    \one_{2\times2} & 0 & 0 & 0 \\
    0 & 0 & 0 & \one_{2\times2} \\
    0 & 0 & \one_{2\times2} & 0 \\
  \end{bmatrix}
  \,.
\end{align}
A peculiarity of Galilean parity reversal is that it is
absent from $\mathrm{U}(1,1,\mathbb{H})$---in fact,
$\eta P \eta = -P = -(P^\dagger)^{-1}$.
For that matter, Galilean parity reversal is also absent from
$\mathrm{SO}(4,1,\mathbb{R})$,
since a transformation that flips the signs of only three components
of a five-vector has determinant $-1$.
Thus, the Galilei subgroup of the five-dimensional Lorentz group
needs to be extended to include parity reversal.

A consequence of the fact that $\eta P \eta \neq (P^\dagger)^{-1}$
is that the four-component spinors $\chi_1$ and $\eta\chi_2$ are no
longer equivalent if the Galilei group is extended to include parity reversal.
Thus, in terms of the extended Galilei group,
the eight-component spinors are the direct sum of two inequivalent
representations, tightening the analogy to Dirac spinors.

%%%%%%%%%%%%%%%%%%%%%%%%%%%%%%%%%%%%%%%%

\subsubsection{Solutions at five-dimensional rest}

Next next us consider solutions to the five-dimensional Dirac equation (\ref{eqn:dirac})
that are at rest in five-dimensional spacetime.
To be sure, this does not mean the system is a physical rest solution:
it means $p^{1,2,3,4}=0$, and thus $p^0 = \sqrt{2mE_0}$
and accordingly $p^+ = \sqrt{mE_0}$.
The equation for positive-energy rest solutions is:
\begin{align}
  \label{eqn:dirac:rest}
  \begin{bmatrix}
    0 & \sqrt{2mE_0} \\
    \sqrt{2mE_0} & 0
  \end{bmatrix}
  \begin{bmatrix}
    \chi_1(0) \\ \eta\chi_2(0)
  \end{bmatrix}
  =
  \sqrt{2mE_0}
  \begin{bmatrix}
    \chi_1(0) \\ \eta\chi_2(0)
  \end{bmatrix}
  \,,
\end{align}
which is solved when $\chi_1(p) = \eta\chi_2(p)$.
There are thus four independent solutions.
As explained above,
these can be categorized by the eigenvalues of
the commuting operators $J_{12}$ and $J_{34}$.
Since
\begin{align}
  J_{12}
  =
  i \lambda_{12}
  =
  \frac{1}{2}
  \begin{bmatrix}
    \sigma_3 & 0 \\
    0 & \sigma_3
  \end{bmatrix}
  \,,
  \qquad
  \qquad
  J_{34}
  =
  i \lambda_{34}
  =
  \frac{1}{2}
  \begin{bmatrix}
    \sigma_3 & 0 \\
    0 & -\sigma_3
  \end{bmatrix}
  \,,
\end{align}
solutions with one non-zero component of $\chi_1(0)$
are the simultaneous eigenstates.

Next, we need to ensure the spinors are invariant under Galilean parity reversal,
namely when the $P$ matrix in Eq.~(\ref{eqn:parity}) acts on them.
This eliminates two of the solutions,
with the remaining solutions being:
\begin{align}
  \label{eqn:spinor:0}
  u_\uparrow(0)
  =
  N
  \begin{bmatrix}
    1 \\
    0 \\
    1 \\
    0 \\
    1 \\
    0 \\
    1 \\
    ~0~ % padding
  \end{bmatrix}
  \qquad
  \qquad
  u_\downarrow(0)
  =
  N
  \begin{bmatrix}
    0 \\
    1 \\
    0 \\
    1 \\
    0 \\
    1 \\
    0 \\
    ~1~ % padding
  \end{bmatrix}
  \,,
\end{align}
which are respectively spin-up and spin-down in the $xy$
plane\footnote{
  In three spatial dimensions, this is the same as saying they're
  spin-up and spin-down along the $z$ axis.
  In four spatial dimensions, ``spin-up'' means spinning from the
  positive $x$ axis towards the positive $y$ axis,
  and ``spin-down'' means spinning in the opposite direction.
}.
Here, $N$ is a normalization factor (to be set later).

%%%%%%%%%%%%%%%%%%%%%%%%%%%%%%%%%%%%%%%%

\subsubsection{Physical solutions in momentum space}

To obtain physical solutions to the Dirac equation,
we first must boost Eq.~(\ref{eqn:spinor:0}) in the unphysical direction
by a rapidity $y = \frac{1}{2} \log\left(\frac{m}{E_0}\right)$
to set $p^+ = m$,
and subsequently perform a Galilei boost of velocity $\bm{v} = \frac{\bm{p}}{m}$.
The relevant boost along the unphysical direction is:
\begin{align}
  M_{\mathrm{unph}}(y)
  &=
  \exp\left\{
    \frac{y}{2}
    \begin{bmatrix}
      0 & 1 \\
      1 & 0
    \end{bmatrix}
    \right\}
  =
  \begin{bmatrix}
    \cosh\frac{y}{2} & \sinh\frac{y}{2} \\
    \sinh\frac{y}{2} & \cosh\frac{y}{2}
  \end{bmatrix}
  \,.
\end{align}
Next, using Eq.~(\ref{eqn:boost:2x2}),
the relevant Galilei boost is:
\begin{align}
  M_{\mathrm{gal}}(\bm{v})
  =
  \exp\big( \bm{\beta}\cdot\bm{v} \big)
  =
  1
  +
  \frac{i}{2\sqrt{2}m}
  \begin{bmatrix*}[r]
    \bm{p}\cdot\bm{\sigma} & \bm{p}\cdot\bm{\sigma} \\
    -\bm{p}\cdot\bm{\sigma} & -\bm{p}\cdot\bm{\sigma} \\
  \end{bmatrix*}
  \,,
\end{align}
where the nilpotency of the generators
(i.e., $\beta_a \beta_b = 0$)
makes the expansion especially simple.

Applying the sequence of boosts $M_{\mathrm{gal}}(\bm{v})M_{\mathrm{unph}}(y)$ to the rest spinor
(\ref{eqn:spinor:0}) according to the transformation law (\ref{eqn:spinor:trans})
gives the physical spinor solutions in momentum space:
\begin{align}
  \label{eqn:spinor}
  u_\uparrow(\bm{p})
  =
  \frac{1}{2^{5/4}m}
  \begin{bmatrix}
    \sqrt{2}m + i p_z \\
    i p_x - p_y \\
    \sqrt{2}m - i p_z \\
    - i p_x + p_y \\
    \sqrt{2mE_0} \\
    0 \\
    \sqrt{2mE_0} \\
    0
  \end{bmatrix}
  \qquad
  \qquad
  u_\downarrow(\bm{p})
  =
  \frac{1}{2^{5/4}m}
  \begin{bmatrix}
    i p_x + p_y \\
    \sqrt{2}m - i p_z \\
    - i p_x - p_y \\
    \sqrt{2}m + i p_z \\
    0 \\
    \sqrt{2mE_0} \\
    0 \\
    \sqrt{2mE_0}
  \end{bmatrix}
  \,,
\end{align}
which are spin-up and spin-down along the $z$ axis, respectively,
and where I have chosen the normalization factor so that
$\bar{u}(\bm{p},s)\gamma^+u(\bm{p},s) = 1$.
Explicit matrix elements for gamma matrices
between these spinors are given in Appendix~\ref{sec:gamma}.
It is curious to note a vague resemblance between these spinors
and the Kogut-Soper spinors~\cite{Kogut:1969xa},
likely due to both sets of spinors being light front spinors
in different spacetimes.

%%%%%%%%%%%%%%%%%%%%%%%%%%%%%%%%%%%%%%%%%%%%%%%%%%%%%%%%%%%%%%%%%%%%%%%%%%%%%%%%

\subsection{Wave equation in coordinate space}

To find the Galilei spinor wave equation in coordinate space,
we must define a coordinate-space wave function.
I use the following definition:
\begin{align}
  \label{eqn:wavefunction}
  \Psi(\bm{x},t)
  \equiv
  \sum_s
  \integral
  \frac{\d^3\bm{p}}{(2\pi)^3}
  \e^{i\bm{p}\cdot\bm{x}}
  u(\bm{p},s)
  \langle \bm{p},s | \Psi(t) \rangle
  \,.
\end{align}
From the momentum-space equation~(\ref{eqn:dirac}),
one can derive the following wave equation:
\begin{align}
  \label{eqn:dirac:space}
  \left(
  i \gamma^+ \frac{\partial}{\partial t}
  +
  i \bm{\gamma}\cdot\bm{\nabla}
  +
  m \gamma^-
  -
  \sqrt{2mE_0}
  \right)
  \Psi(\bm{x},t)
  =
  0
  \,.
\end{align}
This is the coordinate-space wave equation for Galilei spinors.

%%%%%%%%%%%%%%%%%%%%%%%%%%%%%%%%%%%%%%%%

\subsubsection{Two-component spinors}

Since there are only two independent degrees of freedom,
one may ask whether Eq.~(\ref{eqn:dirac:space}) and its solutions
can be recast in terms of two-component objects.
The answer is yes:
with the aid of Eq.~(\ref{eqn:spinor}),
solutions of Eq.~(\ref{eqn:dirac:space}) can be written in terms of
two complex-valued wave functions:
\begin{align}
  \label{eqn:crazy}
  \Psi(\bm{x},t)
  =
  \frac{1}{2^{5/4}m}
  \begin{bmatrix}
    \sqrt{2}m + \partial_z \\
    \partial_x + i\partial_y \\
    \sqrt{2}m - \partial_z \\
    - \partial_x - i\partial_y \\
    \sqrt{2mE_0} \\
    0 \\
    \sqrt{2mE_0} \\
    0
  \end{bmatrix}
  \phi_\uparrow(\bm{x},t)
  +
  \frac{1}{2^{5/4}m}
  \begin{bmatrix}
    \partial_x - i\partial_y \\
    \sqrt{2}m - \partial_z \\
    - \partial_x + i\partial_y \\
    \sqrt{2}m + \partial_z \\
    0 \\
    \sqrt{2mE_0} \\
    0 \\
    \sqrt{2mE_0}
  \end{bmatrix}
  \phi_\downarrow(\bm{x},t)
  \,,
\end{align}
where $\phi_s(\bm{x},t)$ are the Fourier transforms of $\langle \bm{p},s|\Psi(t)\rangle$:
\begin{align}
  \label{eqn:fourier:phi}
  \phi_s(\bm{x},t)
  =
  \integral \frac{\d^3\bm{p}}{(2\pi)^3}
  \e^{i\bm{p}\cdot\bm{x}}
  \langle \bm{p},s | \Psi(t) \rangle
  \,.
\end{align}
The two functions $\phi_s(\bm{x},t)$ can be combined into a two-component object as such:
\begin{align}
  \label{eqn:2comp}
  \psi(\bm{x},t)
  =
  \begin{bmatrix}
    \phi_\uparrow(\bm{x},t)
    \\
    \phi_\downarrow(\bm{x},t)
  \end{bmatrix}
  \,.
\end{align}
This object is conveniently normalized to:
\begin{align}
  \integral \d^3 \bm{x} \,
  \psi^\dagger(\bm{x},t)
  \psi(\bm{x},t)
  =
  1
  \,,
\end{align}
and in the absence of interactions,
its components obey independent Schr\"{o}dinger equations:
\begin{align}
  \label{eqn:schro}
  i \frac{\partial \phi_s(\bm{x},t)}{\partial t}
  =
  E_0
  \phi_s(\bm{x},t)
  -
  \frac{\bm{\nabla}^2}{2m}
  \phi_s(\bm{x},t)
  \,.
\end{align}
This of course seems much easier to deal with than eight-component spinors,
and are accordingly preferable to work with in a fixed frame.
However, the eight-component spinors and their wave equation
(\ref{eqn:dirac:space}) have the correct transformation properties
under Galilei boosts\footnote{
  Technically, it is $\Psi(\bm{x},t) \e^{-i mx^-}$
  that transforms appropriately under Galilei boosts.
  However, since mass is fixed in non-relativistic physics,
  the factor $\e^{-i mx^-}$ can be safely neglected under most circumstances,
  and does not affect observables.
  The factor only needs to be tracked when analyzing transformations of the wave function
  under Galilei boosts, to ensure the phase correctly transforms.
}.
Accordingly, the eight-component spinors are the appropriate
objects to use for constructing currents and matrix elements,
including those appearing in form factor breakdowns.
Additionally, the eight-component spinors contain additional
physical information absent in the two-component formulation,
which is necessary to obtain the correct dynamics.
As an example, I will show as an example
that they entail the correct gyromagnetic ratio $g=2$---a feature also
of the L\'evy-Leblond spinors~\cite{Levy-Leblond:1967eic}.

%%%%%%%%%%%%%%%%%%%%%%%%%%%%%%%%%%%%%%%%

\subsubsection{Gyromagnetic ratio}

One of the most celebrated successes of L\'evy-Leblond spinors
is that they give a gyromagnetic ratio $g=2$,
demonstrating that this is not a strictly relativistic
outcome~\cite{Levy-Leblond:1967eic}.
It is therefore an important sanity check that the eight-component
spinors of this work also give $g=2$.

Electromagnetic interactions are introduced through minimal substitution,
as usual: $\partial_\mu \mapsto \mathcal{D}_\mu = \partial_\mu - i e A_\mu$.
This needs to be done both in the wave equation (\ref{eqn:dirac:space}) itself,
and---if we're building an equation for the two-component spinors---in
the derivatives of Eq.~(\ref{eqn:crazy}).
To introduce a magnetic field in particular,
a time-independent vector potential in particular is introduced.
Performing the required substitutions and working out the rote algebra gives:
\begin{align}
  i \frac{\partial}{\partial t}
  \psi(\bm{x},t)
  =
  \left(
  E_0
  -
  \frac{\big(\bm{\nabla} - i e \bm{A}\big)^2}{2m}
  -
  \frac{e\bm{B}\cdot\bm{\sigma}}{2m}
  \right)
  \psi(\bm{x},t)
  \,,
\end{align}
which is Pauli's equation with $g=2$~\cite{Pauli:1927qhd,Itzykson:1980rh}.

It is worth taking a moment to dwell on this result.
If one were to perform minimal substitution in the
independent Schr\"{o}dinger equations (\ref{eqn:schro}) only,
there would be no interaction with the magnetic field
and we'd find $g=0$;
just as in Pauli's original formulation,
we'd need to put the gyromagnetic factor in by hand.
The superstructure attending the eight-component spinors
contains physics that is otherwise missing from Eq.~(\ref{eqn:schro}),
related to Galilei symmetry---showing that these
eight component spinors are not at all superfluous.

% Explicit matrix elements
\section{Mechanical form factors of spin-half systems}
\label{sec:mff}

Now that we have Galilei-covariant spinors,
we can finally construct a Galilei-covariant expression
for the mechanical form factors.
Like with the spin-zero case,
I write the breakdown in a manner that closely mirrors
existing relativistic breakdowns,
using the expression from Won and Lorc\'{e}~\cite{Won:2025dgc}
in particular as a guide.
The form factor breakdown is:
\begin{multline}
  \label{eqn:emt:half}
  \langle \bm{p}', s' | \hat{T}_q^{\mu\nu}(0) | \bm{p}, s \rangle
  =
  \sqrt{\frac{m}{2E_0}}
  \bar{u}(\bm{p}',s')
  \bigg\{
    \frac{P^\mu P^\nu}{m}
    A_q(\bd^2)
    +
    \frac{\dl^\mu\dl^\nu - g^{\mu\nu}\dl^2}{4m}
    D_q(\bd^2)
    +
    m
    g^{\mu\nu}
    \bar{c}_q(\bd^2)
    +
    P^\mu n^\nu
    \bar{e}_q(\bd^2)
    \\
    +
    \frac{i P^{\{\mu}\sigma^{\nu\}\dl}}{2m}
    J_q(\bd^2)
    -
    \frac{i P^{[\mu} \sigma^{\nu]\dl}}{2m}
    S_q(\bd^2)
    +
    \frac{i E_0 \sigma^{\mu\dl} n^\nu}{2m}
    \bar{f}_q(\bd^2)
    \bigg\}
  u(\bm{p},s)
  \,,
\end{multline}
where $J_q(\bd^2)$ a total angular momentum form factor~\cite{Ji:1996ek}
and $S_q(\bd^2)$ is a fermion spin form factor~\cite{Shore:1999be,Won:2025dgc}.
(The latter can be set to zero if we want to look at the symmetric EMT.)
The form factors $\bar{e}_q(\bd^2)$ and $\bar{f}_q(\bd^2)$ here are new\footnote{
  \added{%
  There is a naming collision with the $\bar{f}_q(\bd^2)$ of
  Polyakov and Sun~\cite{Polyakov:2019lbq}
  for the mechanical form factor breakdown of spin one systems.
  However, the $\bar{f}_q(\bd^2)$ of Polyakov and Sun and the $\bar{f}_q(\bd^2)$
  here are unrelated.%
  }
},
and idiosyncratic to the non-relativistic case;
like $\bar{e}_q(\bd^2)$ in the spin-zero breakdown (\ref{eqn:emt:zero}),
they are a consequence of the fact that energy does not mix into mass or momentum
under boosts, so there can be a component of the energy density that is independent
of the mass and momentum densities.
The overall normalization factor $\sqrt{\frac{m}{2E_0}}$
is introduced to cancel similar factors in the explicit spinor elements
of Eq.~(\ref{eqn:gamma:elements}).

%%%%%%%%%%%%%%%%%%%%%%%%%%%%%%%%%%%%%%%%%%%%%%%%%%%%%%%%%%%%%%%%%%%%%%%%%%%%%%%%

\subsection{Explicit expressions in terms of two-component spinors}
\label{sec:explicit}

The use of eight-component spinors
in Eq.~(\ref{eqn:emt:half})
is necessitated by covariance under
Galilei transformations---including discrete transformations like parity reversal.
However, as shown in Eq.~(\ref{eqn:crazy}),
there are only two independent degrees of freedom for solutions
to the wave equation,
and it is often convenient in practical calculations
to work with two-component spinors (\ref{eqn:2comp}).
To be sure, the latter do not transform under a matrix representation
of the Galilei group~\cite{Huegele:2011},
but within a fixed frame the two-component spinors permit easier
algebraic manipulations.
For this reason, it is to convenient to explicitly evaluate
Eq.~(\ref{eqn:emt:half}) in terms of two-component spinors.
The evaluation is done with the aid of the formulas
in Appendix~\ref{sec:gamma}.

It is easiest to look at collections of components bit-by-bit.
The first few components, relevant to the mass density,
mass flux density, and momentum density respectively, are:
\begin{align}
  \langle \bm{p}', s' | \hat{T}_q^{++}(0) | \bm{p}, s \rangle
  &=
  m
  A_q(\bd^2)
  \delta_{s's}
  \notag
  \\
  \langle \bm{p}', s' | \hat{T}_q^{a+}(0) | \bm{p}, s \rangle
  &=
  P^a
  A_q(\bd^2)
  \delta_{s's}
  -
  \frac{i(\bd\times\bm{\sigma}_{s's})^a}{2}
  \left(
  J_q(\bd^2)
  +
  S_q(\bd^2)
  \right)
  \notag
  \\
  \langle \bm{p}', s' | \hat{T}_q^{+a}(0) | \bm{p}, s \rangle
  &=
  P^a
  A_q(\bd^2)
  \delta_{s's}
  -
  \frac{i(\bd\times\bm{\sigma}_{s's})^a}{2}
  \left(
  J_q(\bd^2)
  -
  S_q(\bd^2)
  \right)
  \,.
\end{align}
The components relevant to the momentum flux density
(i.e., the stress tensor) are:
\begin{multline}
  \langle \bm{p}', s' | \hat{T}_q^{ab}(0) | \bm{p}, s \rangle
  =
  \frac{ P^a P^b }{m}
  A_q(\bd^2)
  \delta_{s's}
  +
  \frac{\dl^a\dl^b - \delta^{ab}\bd^2}{4m}
  D_q(\bd^2)
  \delta_{s's}
  -
  m
  \delta^{ab}
  \bar{c}_q(\bd^2)
  \delta_{s's}
  \\
  -
  \frac{
    i
    (\bd\times\bm{\sigma}_{s's})^{\{a}
    P^{b\}}
  }{2m}
  J_q(\bd^2)
  -
  \frac{
    i
    (\bd\times\bm{\sigma}_{s's})^{[a}
    P^{b]}
  }{2m}
  S_q(\bd^2)
  \,.
\end{multline}
The component relevant to the energy density is:
\begin{multline}
  \label{eqn:half:pm}
  \langle \bm{p}', s' | \hat{T}_q^{+-}(0) | \bm{p}, s \rangle
  =
  \left\{
    \left(
    E_0
    +
    \frac{\bm{P}^2}{2m}
    \right)
    A_q(\bd^2)
    +
    m
    \Big(
    \bar{c}_q(\bd^2)
    +
    \bar{e}_q(\bd^2)
    \Big)
    \right\}
  \delta_{s's}
  \\
  +
  \frac{\bd^2}{4m}
  \left(
  D_q(\bd^2)
  +
  \frac{1}{2}
  A_q(\bd^2)
  -
  J_q(\bd^2)
  +
  S_q(\bd^2)
  \right)
  \delta_{s's}
  -
  \frac{
    i (\bd\times\bm{\sigma}_{s's})\cdot\bm{P}
  }{2m}
  \Big(
  J_q(\bd^2)
  -
  S_q(\bd^2)
  \Big)
  \,.
\end{multline}
Lastly, the components relevant to the energy flux density are:
\begin{multline}
  \langle \bm{p}', s' | \hat{T}_q^{a-}(0) | \bm{p}, s \rangle
  =
  \frac{P^a}{m}
  \left\{
    \left(
    E_0
    +
    \frac{\bm{P}^2}{2m}
    +
    \frac{\bd^2}{8m}
    \right)
    A_q(\bd^2)
    +
    m
    \bar{e}_q(\bd^2)
    -
    \frac{\bd^2}{4m}
    \left(
    J_q(\bd^2)
    -
    S_q(\bd^2)
    \right)
    \right\}
  \delta_{s's}
  \\
  -
  \frac{P^a}{m}
  \frac{
    i (\bd\times\bm{\sigma}_{s's})\cdot\bm{P}
  }{2m}
  \Big(
  J_q(\bd^2)
  -
  S_q(\bd^2)
  \Big)
  \\
  -
  \frac{
    i (\bd\times\bm{\sigma}_{s's})^a
  }{2m}
  \left\{
    E_0
    \Big(
    J_q(\bd^2)
    +
    S_q(\bd^2)
    +
    \bar{f}_q(\bd^2)
    \Big)
    +
    \left(
    \frac{\bm{P}^2}{2m}
    +
    \frac{\bd^2}{8m}
    \right)
    \Big(
    J_q(\bd^2)
    +
    S_q(\bd^2)
    \Big)
    \right\}
  \,.
\end{multline}

%%%%%%%%%%%%%%%%%%%%%%%%%%%%%%%%%%%%%%%%%%%%%%%%%%%%%%%%%%%%%%%%%%%%%%%%%%%%%%%%

\subsection{Density formulas in the Breit frame formalism}
\label{sec:breit}

The standard formalism for defining internal densities of composite systems
is the Breit frame formalism~\cite{Sachs:1962zzc,Halzen:1984mc,Polyakov:2002yz,Polyakov:2018zvc}---although
it is not universally accepted~\cite{Fleming:1974af,Burkardt:2002hr,Miller:2018ybm,Jaffe:2020ebz,Freese:2021czn}.
In Sec.~\ref{sec:pilot},
I will provide yet another competitor to the Breit frame formalism.
However, for the sake of completeness,
I here provide Breit frame density formulas as well.

The relevant formulas are simple to obtain---just
evaluate the matrix element (\ref{eqn:emt:half})
in the $\bm{P}=0$ frame (i.e., the Breit frame),
and perform a Fourier transform over $\bd$:
\begin{align}
  \label{eqn:breit:def}
  \langle T_q^{\mu\nu}(\bm{b}) \rangle_{\text{BF}}
  \equiv
  \sum_{ss'}
  \integral \frac{\d^3\bd}{(2\pi)^3}
  \hat{\psi}^*_{s'} \hat{\psi}_s^{\phantom{*}}
  \langle \bm{p}', s' | \hat{T}_q^{\mu\nu}(0) | \bm{p},s \rangle
  \e^{-i\bd\cdot\bm{b}}
  \bigg|_{\bm{P}=0}
  \,,
\end{align}
where $\hat{\psi}$ is a two-component unit spinor specifying the spin direction:
\begin{align}
  \hat{\psi}
  =
  \begin{bmatrix}
    \e^{-i\phi_s/2}
    \cos\frac{\theta_s}{2}
    \\
    \e^{+i\phi_s/2}
    \sin\frac{\theta_s}{2}
  \end{bmatrix}
  \,,
\end{align}
so that the spin vector is given by:
\begin{align}
  \bs
  =
  \frac{1}{2}
  \hat{\psi}^\dagger \bm{\sigma} \hat{\psi}
  \,.
\end{align}
Then, plugging the explicit matrix elements
at the start of Sec.~\ref{sec:explicit}
into Eq.~(\ref{eqn:breit:def}), the Breit frame densities are:
\begin{align}
  \label{eqn:breit}
  \begin{split}
    \langle T_q^{++}(\bm{b},\bs) \rangle_{\text{BF}}
    &=
    m
    \integral \frac{\d^3\bd}{(2\pi)^3}
    A_q(\bd^2)
    \e^{-i\bd\cdot\bm{b}}
    \\
    \langle T_q^{a+}(\bm{b},\bs) \rangle_{\text{BF}}
    &=
    m
    \integral \frac{\d^3\bd}{(2\pi)^3}
    \frac{\bs\times i\bd}{2m}
    \Big(
    J_q(\bd^2)
    +
    S_q(\bd^2)
    \Big)
    \e^{-i\bd\cdot\bm{b}}
    \\
    \langle T_q^{+a}(\bm{b},\bs) \rangle_{\text{BF}}
    &=
    m
    \integral \frac{\d^3\bd}{(2\pi)^3}
    \frac{\bs\times i\bd}{2m}
    \Big(
    J_q(\bd^2)
    -
    S_q(\bd^2)
    \Big)
    \e^{-i\bd\cdot\bm{b}}
    \\
    \langle T_q^{ab}(\bm{b},\bs) \rangle_{\text{BF}}
    &=
    \integral \frac{\d^3\bd}{(2\pi)^3}
    \left\{
      \frac{\dl^a \dl^b - \delta^{ab}\bd^2}{4m}
      D_q(\bd^2)
      -
      m \delta^{ab}
      \bar{c}_q(\bd^2)
      \right\}
    \e^{-i\bd\cdot\bm{b}}
    \\
    \langle T_q^{+-}(\bm{b},\bs) \rangle_{\text{BF}}
    &=
    \integral \frac{\d^3\bd}{(2\pi)^3}
    \left\{
      E_0
      A_q(\bd^2)
      +
      m
      \big(
      \bar{c}_q(\bd^2)
      +
      \bar{e}_q(\bd^2)
      \big)
      +
      \frac{\bd^2}{4m}
      \left(
      D_q(\bd^2) + \frac{1}{2}A_q(\bd^2) - J_q(\bd^2) + S_q(\bd^2)
      \right)
      \right\}
    \e^{-i\bd\cdot\bm{b}}
    \\
    \langle T_q^{a-}(\bm{b},\bs) \rangle_{\text{BF}}
    &=
    E_0
    \integral \frac{\d^3\bd}{(2\pi)^3}
    \left(1 + \frac{\bd^2}{4mE_0}\right)
    \frac{\bs\times i\bd}{2m}
    \left(
    J_q(\bd^2)
    +
    S_q(\bd^2)
    +
    \bar{f}_q(\bd^2)
    +
    \frac{\bd^2}{4mE_0}
    \Big(
    J_q(\bd^2)
    +
    S_q(\bd^2)
    \Big)
    \right)
    \e^{-i\bd\cdot\bm{b}}
    \,.
  \end{split}
\end{align}

% Spinors in pilot wave theory
\section{Fermion densities in pilot wave theory}
\label{sec:pilot}

With the appropriate Galilei-covariant spinors (\ref{eqn:spinor}) in hand,
we can finally proceed to obtain densities of the energy-momentum tensor.
The extra spin structures in the form factor breakdown complicate
the identification of internal and barycentric contributions to
the mechanical densities,
and furthermore make any ambiguities in the separation more severe.
Like in the spin-zero case, I will use considerations from
the pilot wave formulation of quantum
mechanics~\cite{db:pilot,Bohm:1951xw,Bohm:1951xx,Bohm:2006und}
to guide the separation.
(See Refs.~\cite{Bohm:2006und,holland1995quantum,oriols2019overviewbohmianmechanics}
for excellent introductions to the pilot wave formulation.)

%%%%%%%%%%%%%%%%%%%%%%%%%%%%%%%%%%%%%%%%%%%%%%%%%%%%%%%%%%%%%%%%%%%%%%%%%%%%%%%%

\subsection{Pilot waves for fermions}

Two of the essential ingredients of
the pilot wave formulation of quantum mechanics are
the polar decomposition of the wave function
and the continuity equation.
First, for spin-half particles,
the appropriate polar decomposition can be written
in terms of the two-component spinors (\ref{eqn:2comp})
as~\cite{bohm1955causal,Bohm:2006und,Drezet:2025vcv}:
\begin{align}
  \label{eqn:polar}
  \psi(\bm{x},t)
  =
  \begin{bmatrix}
    \phi_\uparrow(\bm{x},t)
    \\
    \phi_\downarrow(\bm{x},t)
  \end{bmatrix}
  =
  \mathscr{R}(\bm{x},t)
  \e^{i \mathscr{S}(\bm{x},t)}
  \begin{bmatrix}
    \cos\left(
    \frac{\theta_s(\bm{x},t)}{2}
    \right)
    \e^{-i\phi_s(\bm{x},t)/2}
    \\
    \sin\left(
    \frac{\theta_s(\bm{x},t)}{2}
    \right)
    \e^{+i\phi_s(\bm{x},t)/2}
  \end{bmatrix}
  \,.
\end{align}
Here, $\mathscr{R}(\bm{x},t)$ and $\mathscr{S}(\bm{x},t)$ are the real-valued
magnitude and phase functions from the familiar polar decomposition of
spin-zero wave functions,
and $\theta_s(\bm{x},t)$ and $\phi_s(\bm{x},t)$ are angles specifying the spin direction
at the location $\bm{x}$ and time $t$.
The spin at $(\bm{x},t)$ can also be specified by the vector:
\begin{align}
  \label{eqn:spin}
  \bs(\bm{x},t)
  &=
  \frac{1}{2}
  \frac{
    \psi^\dagger(\bm{x},t) \bm{\sigma} \psi(\bm{x},t)
  }{
    \psi^\dagger(\bm{x},t) \psi(\bm{x},t)
  }
  =
  \frac{1}{2}
  \Big(
  \hat{z} \cos\big(\theta_s(\bm{x},t)\big)
  +
  \hat{x} \sin\big(\theta_s(\bm{x},t)\big) \cos\big(\phi_s(\bm{x},t)\big)
  +
  \hat{y} \sin\big(\theta_s(\bm{x},t)\big) \sin\big(\phi_s(\bm{x},t)\big)
  \Big)
  \,.
\end{align}

Second, let us find the continuity equation entailed by the coordinate-space
wave equation (\ref{eqn:dirac:space}).
Defining the conjugate spinor wave
$\overline{\Psi}(\bm{x},t) = \Psi^\dagger(\bm{x},t)\gamma^0$,
solutions to Eq.~(\ref{eqn:dirac:space}) obey the continuity equation:
\begin{align}
  \frac{\partial}{\partial t}
  \Big[
    \overline{\Psi}(\bm{x},t)
    \gamma^+
    \Psi(\bm{x},t)
    \Big]
  +
  \bm{\nabla}\cdot
  \Big[
    \overline{\Psi}(\bm{x},t)
    \bm{\gamma}
    \Psi(\bm{x},t)
    \Big]
  =
  0
  \,.
\end{align}
This suggests the following probability density
and Bohmian velocity:
\begin{align}
  \begin{split}
    \mathscr{P}(\bm{x},t)
    &=
    \overline{\Psi}(\bm{x},t)
    \gamma^+
    \Psi(\bm{x},t)
    \,,
    \\
    \bm{v}_{\bohm}(\bm{x},t)
    &=
    \frac{
      \overline{\Psi}(\bm{x},t)
      \bm{\gamma}
      \Psi(\bm{x},t)
    }{
      \mathscr{P}(\bm{x},t)
    }
    \,.
  \end{split}
\end{align}
With the aid of the matrix elements in Appendix~\ref{sec:gamma},
these can be written in terms of the two-component spinors (\ref{eqn:2comp}) as:
\begin{align}
  \begin{split}
    \mathscr{P}(\bm{x},t)
    &=
    \psi^\dagger(\bm{x},t)
    \psi(\bm{x},t)
    \,,
    \\
    \bm{v}_{\bohm}(\bm{x},t)
    &=
    -\frac{i}{2m}
    \frac{
      \psi^\dagger(\bm{x},t)
      \blrn
      \psi(\bm{x},t)
    }{
      \mathscr{P}(\bm{x},t)
    }
    +
    \frac{
      \bm{\nabla} \times
      \Big(
      \psi^\dagger(\bm{x},t)
      \bm{\sigma}
      \psi(\bm{x},t)
      \Big)
    }{
      2m
      \mathscr{P}(\bm{x},t)
    }
    \,,
  \end{split}
\end{align}
Notably,
I recover Bohm and Hiley's velocity law for spinors~\cite{Bohm:2006und}.
Although it was originally derived through non-relativistic reduction
of Dirac's equation, working in a fully Galilei-covariant framework
is sufficient to obtain the same result.
In fact, Wilkes earlier found that using L\'evy-Leblond spinors
also gives the same velocity law~\cite{Wilkes:2020}.

It is helpful to break the Bohmian velocity down into pieces corresponding to
the convective (or irrotational)
and magnetization (or solenoidal) parts of the probability current:
\begin{align}
  \label{eqn:velocity}
  \begin{split}
    \bm{v}_{\bohm}(\bm{x},t)
    &=
    \bm{v}_{\conv}(\bm{x},t)
    +
    \bm{v}_{\slnd}(\bm{x},t)
    \\
    \bm{v}_{\conv}(\bm{x},t)
    &=
    -\frac{i}{2m}
    \frac{
      \psi^\dagger(\bm{x},t)
      \blrn
      \psi(\bm{x},t)
    }{
      \mathscr{P}(\bm{x},t)
    }
    \\
    \bm{v}_{\slnd}(\bm{x},t)
    &=
    \frac{
      \bm{\nabla} \times
      \Big(
      \psi^\dagger(\bm{x},t)
      \bm{\sigma}
      \psi(\bm{x},t)
      \Big)
    }{
      2m
      \mathscr{P}(\bm{x},t)
    }
    \,.
  \end{split}
\end{align}
Although the currents
$\mathscr{P}\bm{v}_{\conv}$
and
$\mathscr{P}\bm{v}_{\slnd}$
are respectively irrotational and solenoidal
(i.e., they respectively have zero curl and zero divergence),
the velocities themselves may not be.
In fact~\cite{Drezet:2025vcv}:
\begin{align}
  \bm{v}_{\conv}(\bm{x},t)
  &=
  \frac{\bm{\nabla}\mathscr{S}(\bm{x},t)}{m}
  -
  \frac{\cos\big(\theta_s(\bm{x},t)\big)}{2m} \bm{\nabla}\phi_s(\bm{x},t)
  \notag
  \\
  \bm{v}_{\slnd}(\bm{x},t)
  &=
  \frac{
    \bm{\nabla}\mathscr{P}(\bm{x},t) \times \bs(\bm{x},t)
  }{m \mathscr{P}(\bm{x},t)}
  +
  \frac{\bm{\nabla}\times\bs(\bm{x},t)}{m}
  \,.
\end{align}
It is only when the spin direction $\bs$ is fixed that $\bm{v}_{\conv}$
and $\bm{v}_{\slnd}$ respectively have zero curl and zero divergence.

In typical circumstances, the spin direction of a wave function is not fixed.
Fermions interact with their environment in a way that depends on their spin direction,
especially when magnetic fields are present.
Accordingly, the components $\phi_\uparrow(\bm{x},t)$ and $\phi_\downarrow(\bm{x},t)$
of the two-component spinor (\ref{eqn:2comp})
will typically be evolve to be different functions of $(\bm{x},t)$.
On the other hand,
one of the underlying premises of the pilot wave formulation
is that the fermion has a definite position $\bm{x}_{\text{true}}$,
of which we are merely ignorant.
It follows, then, that the fermion also has a definite spin given by
$\bs(\bm{x}_{\text{true}},t)$---similar to how
$\bm{v}_{\bohm}(\bm{x}_{\text{true}},t)$ is postulated to give the actual
velocity\footnote{
  To be sure, like the Bohmian velocity,
  the Bohmian spin $\bs$ is not an observable, but a hidden variable.
  It is distinct from the spin operators $\frac{1}{2}\bm{\sigma}$.
  For instance, a fermion prepared in a spin-up state along the $z$ axis has
  $\varSigma_x = \varSigma_y = 0$,
  while the eigenvalues of $\frac{1}{2}\sigma_x$ and $\frac{1}{2}\sigma_y$ are always $\pm\frac{1}{2}$.
  When a fermion is not already in an eigenstate of a spin projection operator,
  a ``measurement'' of this spin projection disturbs the system and changes the system's spin.
  In fact, one of the perks of the pilot wave formulation is that it provides
  a detailed mechanistic picture of \emph{how} the system is disturbed
  by a measurement; see
  Refs.~\cite{Bell:1982xg,Bohm:2006und,Daumer:1996ca,Norsen:2014,oriols2019overviewbohmianmechanics}
  for more details.
  It should be remembered that in the pilot wave formulation,
  a ``measurement'' of any observable other than position does not
  passively reveal pre-existing properties.
  Accordingly, attributing ontological reality to what
  I have called the Bohmian spin $\bs$ is not in contradiction with
  Bell-type inequalities~\cite{Bell:1964kc,Clauser:1969ny,Bell:1975uz},
  the Kochen-Specker theorem~\cite{Bell:1964fg,Kochen:1968zz}
  or the GHZ experiment~\cite{Mermin:1990,GHZ:1990}.
}.
Any point-to-point variation in the spin direction associated with the wave packet
is a property of the wave packet,
rather than an internal property of the fermion.
Since the goal of this work is to characterize the \emph{internal} densities of fermions,
it is safe to limit our attention to wave packets with a fixed spin direction,
which I will do for the remainder of the paper.

%%%%%%%%%%%%%%%%%%%%%%%%%%%%%%%%%%%%%%%%%%%%%%%%%%%%%%%%%%%%%%%%%%%%%%%%%%%%%%%%

\subsection{From form factors to densities}

To relate the mechanical form factors to densities,
I closely follow derivations given in
Refs.~\cite{Li:2022ldb,Freese:2022fat}.
The expectation value of the EMT for a physical state $|\Psi(t)\rangle$
can be written with the help of completeness relations as:
\begin{align}
  \langle \Psi(t) |
  \hat{T}^{\mu\nu}(\bm{x})
  | \Psi(t) \rangle
  &=
  \sum_{s,s'}
  \integral \frac{\d^3\bm{p}}{(2\pi)^3}
  \integral \frac{\d^3\bm{p}'}{(2\pi)^3}
  \langle \Psi(t) | \bm{p}', s' \rangle
  \langle \bm{p}', s' | \hat{T}^{\mu\nu}(0) | \bm{p}, s \rangle
  \langle \bm{p}, s | \Psi(t) \rangle
  \e^{-i\bd\cdot\bm{x}}
  \,,
\end{align}
where $\bd = \bm{p}' - \bm{p}$, and
where the matrix elements can be written in terms of mechanical form factors
though Eq~(\ref{eqn:emt:half}).
Inverting the Fourier transform (\ref{eqn:fourier:phi}) allows us to write:
\begin{align}
  \langle \Psi(t) |
  \hat{T}^{\mu\nu}(\bm{x})
  | \Psi(t) \rangle
  &=
  \sum_{s,s'}
  \integral \frac{\d^3\bm{p}}{(2\pi)^3}
  \integral \frac{\d^3\bm{p}'}{(2\pi)^3}
  \integral \d^3\bm{r}
  \integral \d^3\bm{r}'
  \,
  \phi^*_{s'}(\bm{r}', t)
  \phi_{s}(\bm{r}, t)
  \langle \bm{p}', s' | \hat{T}^{\mu\nu}(0) | \bm{p}, s \rangle
  \e^{-i(\bd\cdot\bm{x} + \bm{p}\cdot\bm{r} - \bm{p}'\cdot\bm{r}')}
  \,.
\end{align}
Defining the variable transformations:
\begin{align*}
  &
  \bm{P}
  = \frac{1}{2} \big( \bm{p} + \bm{p}' \big)
  \qquad \qquad
  \bd = \bm{p}' - \bm{p}
  \\
  &
  \bm{R}
  = \frac{1}{2} \big( \bm{r} + \bm{r}' \big)
  \qquad \qquad
  \bm{\rho} = \bm{r}' - \bm{r}
  \,,
\end{align*}
we can write:
\begin{align}
  \langle \Psi(t) |
  \hat{T}^{\mu\nu}(\bm{x})
  | \Psi(t) \rangle
  &=
  \sum_{s,s'}
  \integral \frac{\d^3\bm{P}}{(2\pi)^3}
  \integral \frac{\d^3\bd}{(2\pi)^3}
  \integral \d^3\bm{R}
  \integral \d^3\bm{\rho}
  \,
  \phi^*_{s'}(\bm{r}', t)
  \phi_{s}(\bm{r}, t)
  \langle \bm{p}', s' | \hat{T}^{\mu\nu}(0) | \bm{p}, s \rangle
  \e^{-i\bd\cdot(\bm{x}-\bm{R})}
  \e^{i\bm{P}\cdot\bm{\rho}}
  \,.
\end{align}
Just as in Refs.~\cite{Li:2022ldb,Freese:2022fat},
the integral over $\bm{P}$ can be performed if all instances of $\bm{P}$
appearing in the matrix element
$\langle \bm{p}', s' | \hat{T}^{\mu\nu}(0) | \bm{p}, s \rangle$
are replaced by
the two-sided derivative $-\frac{i}{2} \blrn$,
acting between
$\phi^*_{s'}(\bm{r}', t)$
and
$\phi_{s}(\bm{r}, t)$.
The integral over $\bm{P}$ then creates a delta function setting $\bm{\rho}=0$,
giving the expression:
\begin{align}
  \label{eqn:emt:exp}
  \langle \Psi(t) | \hat{T}^{\mu\nu}(\bm{x}) | \Psi(t) \rangle
  &=
  \sum_{s,s'}
  \integral \d^3\bm{R}
  \integral \frac{\d^3\bd}{(2\pi)^3}
  \phi^*_{s'}(\bm{R}, t)
  \langle \bm{p}', s' | \hat{T}^{\mu\nu}(0) | \bm{p}, s \rangle
  \bigg|_{2i\bm{P}\rightarrow\blrn}
  \phi_{s}(\bm{R}, t)
  \e^{-i\bd\cdot(\bm{x}-\bm{R})}
  \,.
\end{align}
At this point, the formulas in Sec.~\ref{sec:explicit} can be used
to evaluate the densities for fermions prepared in realistic wave packets.

%%%%%%%%%%%%%%%%%%%%%%%%%%%%%%%%%%%%%%%%%%%%%%%%%%%%%%%%%%%%%%%%%%%%%%%%%%%%%%%%

\subsection{Densities of elementary fermions}

I will now use Eq.~(\ref{eqn:emt:exp}) to evaluate densities of elementary fermions.
Elementary fermions are a helpful intuition-building exercise,
since they should have minimal internal structure, and accordingly
most of the mass, momentum and energy distributions should be attributable
to the wave packet.
There are actually two cases to consider:
(1) the asymmetric EMT, for which $S(\bd^2) = \frac{1}{2}$; and
(2) the symmetric EMT, for which $S(\bd^2) = 0$.
In both cases, the remaining form factors are given by
$A(\bd^2) = 1$,
$J(\bd^2) = \frac{1}{2}$,
and $D(\bd^2) = \bar{c}(\bd^2) = \bar{e}(\bd^2) = \bar{f}(\bd^2) = 0$\footnote{
  See Ref.~\cite{Hudson:2017oul} regarding the zero $D(\bd^2)$.
  The zero $\bar{c}(\bd^2)$ and $\bar{e}(\bd^2)$ are related to
  sum rules; cf.\ Sec~\ref{sec:sum}.
  The zero $\bar{f}_q(\bd^2)$ is effectively a statement that
  mass and energy don't flow differently in an elementary fermion.
}.

I will consider the mass density first.
In either case, it is simply given by:
\begin{align}
  \langle \Psi(t) | \hat{T}^{++}(\bm{x}) | \Psi(t) \rangle
  =
  m
  \mathscr{P}(\bm{x},t)
  \,,
\end{align}
which is compatible with the notion of the fermion being pointlike.
The density in this case is entirely due to
the uncertainty in the fermion's position.

Next, I consider the mass flux density.
The result differs depending on whether the asymmetric or symmetric EMT is used.
For the asymmetric EMT:
\begin{align}
  \label{eqn:a+:asym}
  \langle \Psi(t) | \hat{T}_{\text{asym}}^{a+}(\bm{x}) | \Psi(t) \rangle
  =
  m
  \mathscr{P}(\bm{x},t)
  v_{\bohm}^a(\bm{x},t)
  \,,
\end{align}
which is compatible with the notion of a pointlike particle
moving with the Bohmian velocity given by Eq.~(\ref{eqn:velocity}).
On the other hand, the symmetric EMT gives a mass flux density of:
\begin{align}
  \langle \Psi(t) | \hat{T}_{\text{sym}}^{a+}(\bm{x}) | \Psi(t) \rangle
  =
  m
  \mathscr{P}(\bm{x},t)
  \left(
  v_{\conv}^a(\bm{x},t)
  +
  \frac{1}{2}
  v_{\slnd}^a(\bm{x},t)
  \right)
  \,,
\end{align}
where $\bm{v}_{\conv}$ and $\bm{v}_{\slnd}$ are defined in
Eq.~(\ref{eqn:velocity}).
The interpretation of this expression is unclear,
and appears at odds with the notion of a pointlike particle
moving according to the Bohmian velocity law~(\ref{eqn:velocity}).
The pilot wave interpretation accordingly seems to favor the asymmetric EMT.

Next, I consider the momentum density.
This again differs for the asymmetric and symmetric EMT.
In the first case:
\begin{align}
  \langle \Psi(t) | \hat{T}_{\text{asym}}^{+a}(\bm{x}) | \Psi(t) \rangle
  =
  m
  \mathscr{P}(\bm{x},t)
  v_{\conv}^a(\bm{x},t)
  \equiv
  \mathscr{P}(\bm{x},t)
  p_{\bohm}^a(\bm{x},t)
  \,,
\end{align}
which in effect defines the Bohmian momentum.
In fact, this expression for the Bohmian momentum agrees exactly
with the result of Bohm and Hiley~\cite{Bohm:2006und},
which itself follows from the quantum Hamilton-Jacobi equation~\cite{bohm1955causal}.
Accordingly, I also reproduce their observation that:
\begin{align}
  \bm{p}_{\bohm}(\bm{x},t)
  \neq
  m \bm{v}_{\bohm}(\bm{x},t)
  \,,
\end{align}
which, to be sure, is a generic feature of having an asymmetric EMT
(see the discussion by Hehl~\cite{Hehl:1976vr} for instance).
This again suggests that the pilot wave interpretation prefers the asymmetric EMT;
the symmetric EMT instead gives equal momentum and mass flux densities,
both of which involve the peculiar mixture $\bm{v}_{\conv} + \frac{1}{2} \bm{v}_{\slnd}$.

Since the pilot wave interpretation favors it,
I will focus exclusively on the asymmetric EMT for the remainder of this work.
The inequivalence between momentum and velocity however raises a pertinent question:
how do we define the rest frame in the pilot wave formulation?
Should it be the zero-velocity frame, or the zero-momentum frame?
Each choice will lead to different conclusions about the fermion's internal densities,
since boosts from the rest frame mix components of the EMT---see
Eq.~(\ref{eqn:full:zero}).

The zero-momentum frame seems like the natural choice for the rest frame.
Although the Bohmian velocity is non-zero in this frame,
being given by $\bm{v}_{\slnd}$ (see Eq.~(\ref{eqn:velocity})),
it can be interpreted as revolutionary motion about some pivot~\cite{Bohm:2006und},
where the pivot itself moves with the convective velocity $\bm{v}_\nabla$.
On the other hand, in the zero-velocity frame,
there is non-zero revolutionary momentum \emph{opposite} the spin direction,
which has no sensible physical interpretation.

Internal densities of the fermion are defined relative to
the rest frame of the pivot---i.e.,
relative to the zero-momentum frame.
Accordingly, motion relative to the pivot should be considered
part of the fermion's internal structure.
This includes the mass flux density.
For an elementary fermion with a spin vector $\bs$,
the internal mass flux density is thus:
\begin{align}
  \blt_{\elem}^{a+}(\bm{b},\bs)
  =
  -
  (\bs\times\bm{\nabla})^a
  \delta^{(3)}(\bm{b})
  \,.
\end{align}
Due to our ignorance of the pivot's exact location,
we must smear this out by the probability density,
meaning the smeared contribution is:
\begin{align}
  \integral \d^3 \bm{R} \,
  \psi^\dagger(\bm{R},t) \psi(\bm{R},t)
  \,
  \blt^{a+}\big(\bm{x}-\bm{R}, \bs(\bm{R},t)\big)
  =
  \frac{1}{2}
   \Big(
   \bm{\nabla} \times
   \big(
   \psi^\dagger(\bm{x},t)
   \bm{\sigma}
   \psi(\bm{x},t)
   \big)
   \Big)^a
  \,.
\end{align}
Subsequently boosting this by the pivot velocity gives Eq.~(\ref{eqn:a+:asym}).

The remaining components of the energy-momentum tensor
can be broken down in a similar manner to Eq.~(\ref{eqn:full:zero}),
albeit using the pivot velocity $\bm{v}_{\conv}$
in the Galilei boost matrix (\ref{eqn:boost}).
The question now arises of how exactly to separate the rest-frame
EMT into an intrinsic part $\blt^{\alpha\beta}$ and
a quantum part $T_Q^{\alpha\beta}$.
It is here that considering the case of fixed spin is helpful,
for two reasons.
First, any point-to-point variation of of the spin vector $\bs$ is a property
of the wave packet, rather than of the fermion's internal structure;
thus any terms that are dropped by considering a constant spin vector
are part of $T_Q^{\alpha\beta}$.
Second, when the spin vector is constant,
the quantum EMT is identical to the spin-zero case (\ref{eqn:emt:quantum}).
The presence of point-to-point variations would introduce
additional terms to the quantum potential~\cite{bohm1955causal,Bohm:2006und},
and additional quantum stresses corresponding to torques
that rotate the spin vector~\cite{bohm1955causal},
but these matters are beyond the scope of this work
and need not be considered to study internal structure.

With the spin vector now fixed, the procedure is simple.
First, we evaluate Eq.~(\ref{eqn:emt:exp}) using the explicit
matrix elements in Sec.~\ref{sec:explicit}.
Second, we perform the \emph{inverse} Galilei boosts on the integrand
of the result---i.e., the integrand is boosted by $-\bm{v}_{\conv}$.
Third, in accordance with Eq.~(\ref{eqn:full:zero}),
the quantum EMT (\ref{eqn:emt:quantum}) is subtracted off.
What remains is the internal EMT of an elementary fermion,
smeared by the probability density $\mathscr{P}$.
The resulting non-zero components of the internal EMT are:
\begin{align}
  \label{eqn:elementary}
  \begin{split}
    &
    \blt_{\elem}^{++}(\bm{b},\bs)
    =
    m \delta^{(3)}(\bm{b})
    \qquad
    \qquad
    \blt_{\elem}^{a+}(\bm{b},\bs)
    =
    -
    (\bs\times\bm{\nabla})^a
    \delta^{(3)}(\bm{b})
    \\
    &
    \blt_{\elem}^{+-}(\bm{b},\bs)
    =
    E_0 \delta^{(3)}(\bm{b})
    \qquad
    \qquad
    \blt_{\elem}^{a-}(\bm{b},\bs)
    =
    -
    \frac{E_0}{m}
    (\bs\times\bm{\nabla})^a
    \delta^{(3)}(\bm{b})
    \,.
  \end{split}
\end{align}
In essence, the elementary fermion carries only mass and energy in
the zero-momentum frame,
which are rapidly revolving around a pivot that is at rest.

%%%%%%%%%%%%%%%%%%%%%%%%%%%%%%%%%%%%%%%%%%%%%%%%%%%%%%%%%%%%%%%%%%%%%%%%%%%%%%%%

\subsection{Composite fermions}

Finally, let's look at densities of composite fermions.
Most of the conceptual and formal elements needed to
isolate the internal densities are now in place.
The missing ingredient is an appropriate generalization of the
breakdown~(\ref{eqn:full:zero}).
We could try to apply it in its current form to the special case
of a constant spin vector $\bs$,
but we would find an extra term in the energy flux density
that cannot be sorted into either of the terms present in
Eq.~(\ref{eqn:full:zero}).
Nonetheless, I will proceed and patch up Eq.~(\ref{eqn:full:zero}) afterwards.

The method is basically the same as we used to obtain Eq.~(\ref{eqn:elementary}).
We first work out
Eq.~(\ref{eqn:emt:exp}) using the matrix elements in Sec.~\ref{sec:explicit},
and then apply the \emph{inverse} of the Galilei boost matrix (\ref{eqn:boost})
to the integrand.
Since constant $\bs$ is assumed,
Eq.~(\ref{eqn:emt:quantum}) gives the quantum EMT,
which is then subtracted off from the result,
under the assumption that Eq.~(\ref{eqn:full:zero}) is applicable.
Ideally, the remaining terms should be equal to a convolution between $\mathscr{P}$
and a tensor that is independent of $\mathscr{P}$---the latter
being interpretable as an internal density.
This almost works out,
but there is one extra term in the energy flux density that violates this expectation.
Putting this extra term aside,
the internal densities suggested by Eq.~(\ref{eqn:full:zero}) are:
\begin{align}
  \label{eqn:main}
  \begin{split}
    \blt_q^{++}(\bm{b},\bs)
    &=
    m
    \integral \frac{\d^3\bd}{(2\pi)^3}
    A_q(\bd^2)
    \e^{-i\bd\cdot\bm{b}}
    \equiv
    m
    \bla_q(\bm{b})
    \\
    \blt_q^{a+}(\bm{b},\bs)
    &=
    m
    \integral \frac{\d^3\bd}{(2\pi)^3}
    \frac{\bs\times i\bd}{2m}
    \Big(
    J_q(\bd^2)
    +
    S_q(\bd^2)
    \Big)
    \e^{-i\bd\cdot\bm{b}}
    \\
    \blt_q^{+a}(\bm{b},\bs)
    &=
    m
    \integral \frac{\d^3\bd}{(2\pi)^3}
    \frac{\bs\times i\bd}{2m}
    \Big(
    J_q(\bd^2)
    -
    S_q(\bd^2)
    \Big)
    \e^{-i\bd\cdot\bm{b}}
    \\
    \blt_q^{ab}(\bm{b},\bs)
    &=
    \integral \frac{\d^3\bd}{(2\pi)^3}
    \left\{
      \frac{\dl^a \dl^b - \delta^{ab}\bd^2}{4m}
      D_q(\bd^2)
      -
      m \delta^{ab}
      \bar{c}_q(\bd^2)
      \right\}
    \e^{-i\bd\cdot\bm{b}}
    \\
    \blt_q^{+-}(\bm{b},\bs)
    &=
    \integral \frac{\d^3\bd}{(2\pi)^3}
    \left\{
      E_0
      A_q(\bd^2)
      +
      m
      \big(
      \bar{c}_q(\bd^2)
      +
      \bar{e}_q(\bd^2)
      \big)
      +
      \frac{\bd^2}{4m}
      \Big(
      D_q(\bd^2) - J_q(\bd^2) + S_q(\bd^2)
      \Big)
      \right\}
    \e^{-i\bd\cdot\bm{b}}
    \\
    \blt_q^{a-}(\bm{b},\bs)
    &=
    E_0
    \integral \frac{\d^3\bd}{(2\pi)^3}
    \frac{\bs\times i\bd}{2m}
    \Big(
    J_q(\bd^2)
    +
    S_q(\bd^2)
    +
    \bar{f}_q(\bd^2)
    \Big)
    \e^{-i\bd\cdot\bm{b}}
    \,.
  \end{split}
\end{align}
In fact, once we have properly accounted for the extra term,
these turn out to be the correct internal densities.
This pesky extra term in the energy flux density can be written:
\begin{align*}
  \delta^\nu_-
  \integral \d^3 \bm{R} \,
  T_Q^{\mu b}(\bm{R},t)
  \blt^{+b}(\bm{x}-\bm{R},\bs)
  \,.
\end{align*}
It is apparent that this term is \emph{not} of the required form,
i.e., it is not $\mathscr{P}$ convolved with a $\mathscr{P}$-independent function.
Nonetheless, the contraction of the internal momentum density
with the quantum stress tensor suggests a possible physical origin:
the quantum EMT must be boosted by the \emph{total} convective velocity
of the matter that the wave function is acting on,
rather than just the convective velocity of the barycenter.
This suggests the following generalization of Eq.~(\ref{eqn:full:zero}):
\begin{multline}
  \label{eqn:full:half}
  \langle \Psi(t) | \hat{T}_q^{\mu\nu}(\bm{x}) | \Psi(t) \rangle
  =
  \integral \d^3 \bm{R} \,
  \bigg\{
    \varLambda^{\mu}_{\phantom{\mu}\alpha}\big(\bm{v}_{\conv}(\bm{R},t)\big)
    \varLambda^{\nu}_{\phantom{\nu}\beta} \big(\bm{v}_{\conv}(\bm{R},t)\big)
    \mathscr{P}(\bm{R},t)
    \blt_q^{\alpha\beta}(\bm{x}-\bm{R},\bs)
    \\
    +
    \varLambda^{\mu}_{\phantom{\mu}\alpha}\Big(\bm{v}_{\conv}(\bm{R},t) + \bm{u}_q(\bm{x}-\bm{R},\bs)\Big)
    \varLambda^{\nu}_{\phantom{\nu}\beta} \Big(\bm{v}_{\conv}(\bm{R},t) + \bm{u}_q(\bm{x}-\bm{R},\bs)\Big)
    T_Q^{\alpha\beta}(\bm{R},t)
    \bla_q(\bm{x}-\bm{R})
    \bigg\}
  \,,
\end{multline}
where $\bm{u}_{q}$ is the \emph{internal} convective velocity:
\begin{align}
  \bm{u}_{q}(\bm{b},\bs)
  =
  \frac{
    \blt^{+a}_q(\bm{b},\bs)
  }{
    \blt^{++}_q(\bm{b})
  }
  \,.
\end{align}
Now Eq.~(\ref{eqn:full:zero}) is just a special case of Eq.~(\ref{eqn:full:half}),
since $\bm{u}_{q}$ vanishes for spin-zero states.
Moreover, by explicitly plugging the internal densities (\ref{eqn:main})
and the quantum EMT (\ref{eqn:emt:quantum})
into Eq.~(\ref{eqn:full:half}),
one does arrive at the same result as evaluating Eq.~(\ref{eqn:emt:exp}).
This consistency check validates the formulas in
question---and so Eq.~(\ref{eqn:main}) gives the internal mechanical densities
of composite fermions in the pilot wave formulation.

Before concluding, it is worth taking a moment to compare and contrast
the mechanical densities in the pilot wave formulation (\ref{eqn:main})
and the Breit frame formalism (\ref{eqn:breit}).
Remarkably, for nearly all components, the densities agree---the sole
exceptions being the energy and energy flux densities.
In both of these cases, the pilot wave expressions are simpler.
While this does not prove the correctness of the pilot wave picture,
it does lend some credence to the idea that Breit frame densities are
contaminated by artifacts of wave function dispersion.

% Summary/conclusions/outlook
\section{Summary and outlook}
\label{sec:end}

The goals of this work were twofold:
(1) obtain Galilei-covariant expressions for matrix elements of the energy-momentum tensor; and
(2) obtain formulas for internal energy-momentum densities of non-relativistic composite systems.
The first objective was achieved in
Eq.~(\ref{eqn:emt:zero})
for spin-zero systems, and
Eq.~(\ref{eqn:emt:half})
for spin half systems.
For the spin-half expression in particular,
a lengthy detour into group theory was necessary to build a new set of
spinors---given by Eq.~(\ref{eqn:spinor})---that respect all the necessary symmetries
(namely, the Galilei group extended by parity).

The second objective was addressed in two formalisms:
the standard Breit frame formalism,
where one simply sets $\bm{P}=0$ in the matrix element
and takes the Fourier transform;
and the pilot wave formalism,
where the composite particle is assumed to have a definite position,
momentum and energy at all times,
and where contributions from the barycentric wave packet
can simply be subtracted off.
The Breit frame densities are given in Eq.~(\ref{eqn:breit}),
and the pilot wave densities in Eq.~(\ref{eqn:main}).
Most of these densities are identical, with the exceptions being
the energy density and energy flux density.
This discrepancy occurs because part of the Breit frame matrix element---namely,
the $\frac{\bd^2}{8m} A_q(\bd^2)$ term in the case of
the energy density---is attributed by the pilot wave formalism
to wave packet dispersion.

Put another way:
the Breit frame Fourier transform involves matrix elements between plane wave states
with non-zero momentum---see Eq.~(\ref{eqn:breit:def})---meaning that
each of these plane waves would have a barycentric kinetic energy $\frac{\bd^2}{8m}$.
The Breit frame formalism in effect considers this barycentric energy
to be part of the ``internal'' energy density.
It thus appears that the Breit frame energy density is contaminated by
what is actually part of the barycentric energy density.
This contamination is absent in the pilot wave result.

This also raises the question of whether---and to what degree---expressions
for relativistic densities are also contaminated by artifacts of wave packet dispersion.
Most of the literature critical of relativistic Breit frame densities
is concerned with relativistic effects associated with Lorentz
boosts~\cite{Burkardt:2002hr,Miller:2018ybm,Jaffe:2020ebz,Freese:2021czn,Freese:2022fat},
while the concern I raise here is present even with Galilei symmetry.
In principle, existing expressions for light front energy
densities (e.g.\ in Ref.~\cite{Freese:2022fat})
may be contaminated by wave packet dispersion,
just as much as Breit frame densities are.

It thus seems pertinent to ask whether
the pilot wave based analysis of Sec.~\ref{sec:pilot}
can be repeated for relativistic Dirac spinors,
to aid in the removal of such contamination.
This will be the main subject of a follow-up study.
The pilot wave formulation infamously has difficulties with
Lorentz covariance when applied to relativistic
systems~\cite{Bohm:2006und,Berndl:1995ne,Duerr:1997ink,Durr:2013asa,Struyve:2024onb},
although significant progress has been made in the Bohmian formulation of Dirac
fields, including a Lorentz-covariant polar
decomposition~\cite{Fabbri:2020ypd,Fabbri:2022kfr,Fabbri:2023onb}.
Such an analysis would also be more directly relevant to
mechanical properties of the proton in particular,
and would warrant concrete numerical and visual demonstrations
in addition to formulas.

In the meantime,
the most likely application of the results in this work
will be to spin-zero and spin-one bound states \emph{composed of}
non-relativistic fermions.
This includes calculating and interpreting mechanical properties of
heavy quarkonia in non-relativistic quantum chromodynamics
(NRQCD)~\cite{Lucha:1991vn,Brambilla:1999xf}.
While the form factor breakdown of these systems will of course differ
from Eq.~(\ref{eqn:emt:half}),
this formula will still be relevant:
the effective EMT operator of the spin-half constituents
will need to be constructed
to reproduce Eq.~(\ref{eqn:emt:half}) when sandwiched between
one-particle states.

%%%%%%%%%%%%%%%%%%%%%%%%%%%%%%%%%%%%%%%%

\begin{acknowledgments}
  This work benefited from discussions at the workshop
  \textsl{Mechanical properties of hadrons: Structure, dynamics, visualization}
  at the
  European Centre for Theoretical Studies in Nuclear Physics and Related Areas (ECT*-FBK),
  especially from pertinent conversations with
  Enrique Ruiz Arriola, Jambul Gegelia, C\'edric Lorc\'e and Peter Schweitzer.
  I'm also grateful for valuable discussions with
  Daniel Adamiak, Wim Cosyn, Jian-Wei Qiu, Ted Rogers, Gabriel Santiago, Alan Sosa, Frank Vera and Christian Weiss.
  The research in this work received inspiration from the goals of the
  Quark Gluon Tomography Topical Collaboration of the U.S.\ Department of Energy.
  This work was supported by the DOE contract No.~DE-AC05-06OR23177,
  under which Jefferson Science Associates, LLC operates Jefferson Lab,
  and by the Scientific Discovery through Advanced Computing (SciDAC) award
  \textsl{Femtoscale Imaging of Nuclei using Exascale Platforms}.
\end{acknowledgments}

\appendix

% Explicit gamma matrices and spinor elements
\section{Explicit five-dimensional gamma matrices and matrix elements}
\label{sec:gamma}

To make this Appendix more helpful and self-contained,
I will reproduce several formulas present in the main text,
in addition to explicit expressions for the gamma matrices
and spinor elements.

The tau matrices are defined in Eq.~(\ref{eqn:vm}),
and reproduced here for convenience:
\begin{align}
  \tau_0
  =
  \begin{bmatrix}
    1 & 0 \\
    0 & 1
  \end{bmatrix}
  \qquad
  %
  %&
  \tau_1
  =
  \begin{bmatrix*}[r]
    0 & -\qi \\
    \qi & 0
  \end{bmatrix*}
  \qquad
  \tau_2
  =
  \begin{bmatrix*}[r]
    0 & -\qj \\
    \qj & 0
  \end{bmatrix*}
  \qquad
  \tau_3
  =
  \begin{bmatrix*}[r]
    0 & -\qk \\
    \qk & 0
  \end{bmatrix*}
  \qquad
  \tau_4
  =
  \begin{bmatrix}
    0 & 1 \\
    1 & 0
  \end{bmatrix}
  \,.
\end{align}
The gamma matrices are defined in terms of them via:
\begin{align}
  \gamma^\mu
  \equiv
  \begin{bmatrix}
    0 & \tau^\mu \\
    \eta \tau^\mu \eta & 0
  \end{bmatrix}
  \,,
\end{align}
where
\begin{align}
  \eta
  =
  \begin{bmatrix}
    1 & 0 \\
    0 & -1
  \end{bmatrix}
\end{align}
is the ``metric'' for which matrices in $\mathrm{U}(1,1,\mathbb{H})$
are isometries.
The quaternions can be mapped to complex matrices via:
\begin{align}
  \qi
  \equiv
  -i\sigma_1
  \qquad
  \qj
  \equiv
  -i\sigma_2
  \qquad
  \qk
  \equiv
  -i\sigma_3
  \,,
\end{align}
allowing the tau and gamma matrices to be written entirely in terms of complex numbers.
The explicit complex-valued expressions for the gamma matrices are:
\begin{align}
  \begin{split}
    &
    \gamma^0
    =
    \begin{bmatrix}
      0 & 0 & 1 & 0 \\
      0 & 0 & 0 & 1 \\
      1 & 0 & 0 & 0 \\
      0 & 1 & 0 & 0
    \end{bmatrix}
    \qquad
    \gamma^4
    =
    \begin{bmatrix*}[r]
      0 & 0 &  0 & -1 \\
      0 & 0 & -1 &  0 \\
      0 & 1 &  0 &  0 \\
      1 & 0 &  0 &  0
    \end{bmatrix*}
    \qquad
    \gamma^+
    =
    \frac{1}{\sqrt{2}}
    \begin{bmatrix*}[r]
      0 & 0 &  1 & -1 \\
      0 & 0 & -1 &  1 \\
      1 & 1 &  0 &  0 \\
      1 & 1 &  0 &  0
    \end{bmatrix*}
    \qquad
    \gamma^-
    =
    \frac{1}{\sqrt{2}}
    \begin{bmatrix*}[r]
      0 &  0 & 1 & 1 \\
      0 &  0 & 1 & 1 \\
      1 & -1 & 0 & 0 \\
     -1 &  1 & 0 & 0
    \end{bmatrix*}
    \\
    % ~~~~~~~~~~~~~~~~~~~~~~~~~~~~~~~~~~~~
    &
    \gamma^a
    =
    \begin{bmatrix}
      0           & 0          & 0          & -i\sigma_a \\
      0           & 0          & i\sigma_a & 0 \\
      0           & i\sigma_a & 0          & 0 \\
      -i\sigma_a & 0          & 0          & 0
    \end{bmatrix}
    \,,
  \end{split}
\end{align}
for $a\in\{1,2,3\}$.
The sigma matrices,
defined as $\sigma^{\mu\nu} = \frac{i}{2}[\gamma^\mu,\gamma^\nu]$,
are also helpful to have on hand:
\begin{align}
  \begin{split}
    &
    \sigma^{0a}
    =
    \begin{bmatrix}
      0 & -\sigma_a & 0 & 0 \\
      \sigma_a & 0 & 0 & 0 \\
      0 & 0 & 0 & \sigma_a \\
      0 & 0 & -\sigma_a & 0
    \end{bmatrix}
    \qquad
    \sigma^{4a}
    =
    \begin{bmatrix}
      -\sigma_a & 0 & 0 & 0 \\
      0 & \sigma_a & 0 & 0 \\
      0 & 0 & -\sigma_a & 0 \\
      0 & 0 & 0 & \sigma_a
    \end{bmatrix}
    \qquad
    \sigma^{04}
    =
    - \sigma^{+-}
    =
    i
    \begin{bmatrix*}[r]
      0 & 1 & 0 & 0 \\
      1 & 0 & 0 & 0 \\
      0 & 0 & 0 & -1 \\
      0 & 0 & -1 & 0
    \end{bmatrix*}
    \\
    % ~~~~~~~~~~~~~~~~~~~~~~~~~~~~~~~~~~~~
    &
    \sigma^{ab}
    =
    \epsilon_{abc}
    \begin{bmatrix}
      \sigma_c & 0 & 0 & 0 \\
      0 & \sigma_c & 0 & 0 \\
      0 & 0 & \sigma_c & 0 \\
      0 & 0 & 0 & \sigma_c
    \end{bmatrix}
    \qquad
    \sigma^{+a}
    =
    \frac{1}{\sqrt{2}}
    \begin{bmatrix}
      -\sigma_a & -\sigma_a & 0 & 0 \\
      \phantom{-}\sigma_a & \phantom{-}\sigma_a & 0 & 0 \\
      0 & 0 & -\sigma_a & \sigma_a \\
      0 & 0 & -\sigma_a & \sigma_a
    \end{bmatrix}
    \qquad
    \sigma^{-a}
    =
    \frac{1}{\sqrt{2}}
    \begin{bmatrix}
      \sigma_a & -\sigma_a & 0 & 0 \\
      \sigma_a & -\sigma_a & 0 & 0 \\
      0 & 0 & \phantom{-}\sigma_a & \phantom{-}\sigma_a \\
      0 & 0 & -\sigma_a & -\sigma_a
    \end{bmatrix}
    \,.
  \end{split}
\end{align}
For convenience, I reproduce the physical spinor solutions
(\ref{eqn:spinor}) here as well:
\begin{align}
  u_\uparrow(\bm{p})
  =
  \frac{1}{2^{5/4}m}
  \begin{bmatrix}
    \sqrt{2}m + i p_z \\
    i p_x - p_y \\
    \sqrt{2}m - i p_z \\
    - i p_x + p_y \\
    \sqrt{2mE_0} \\
    0 \\
    \sqrt{2mE_0} \\
    0
  \end{bmatrix}
  \qquad
  \qquad
  u_\downarrow(\bm{p})
  =
  \frac{1}{2^{5/4}m}
  \begin{bmatrix}
    i p_x + p_y \\
    \sqrt{2}m - i p_z \\
    - i p_x - p_y \\
    \sqrt{2}m + i p_z \\
    0 \\
    \sqrt{2mE_0} \\
    0 \\
    \sqrt{2mE_0}
  \end{bmatrix}
  \,,
\end{align}
which respectively have spin-up and spin-down
along the $z$ axis (i.e., within the $xy$ plane).

Next, I will give matrix elements of the gamma matrices
between the physical spin-up and spin-down solutions.
The relevant matrix elements are between spinors with
different initial and final five-momenta $p$ and $p'$,
where I also use standard notation for the
average $P = \frac{1}{2}\big(p+p'\big)$,
and momentum transfer $\dl = p'-p$.
We require $p^+ = p'^+ = P^+$ and thus $\dl^+=0$,
since $P^+=m$ is the physical mass
(which does not change in non-relativistic processes).
I will use the notation:
\begin{align}
  \llbracket
  \gamma^\mu
  \rrbracket
  =
  \begin{bmatrix}
    ~ % padding
    \bar{u}_\uparrow(p') \gamma^\mu u_\uparrow(p)
    ~ % padding
    &
    ~ % padding
    \bar{u}_\uparrow(p') \gamma^\mu u_\downarrow(p)
    ~ % padding
    \\
    ~ % padding
    \bar{u}_\downarrow(p') \gamma^\mu u_\uparrow(p)
    ~ % padding
    &
    ~ % padding
    \bar{u}_\downarrow(p') \gamma^\mu u_\downarrow(p)
    ~ % padding
  \end{bmatrix}
\end{align}
to write the results more compactly.
The resulting matrix elements are:
\begin{align}
  \label{eqn:gamma:elements}
  \begin{split}
    \llbracket
    \one
    \rrbracket
    &=
    \sqrt{\frac{2E_0}{m}}
    \delta_{s's}
    \\
    \llbracket
    \gamma^+
    \rrbracket
    &=
    \delta_{s's}
    \\
    \llbracket
    \gamma^a
    \rrbracket
    &=
    \frac{P^a}{m}
    \delta_{s's}
    -
    \frac{i (\bd \times \bm{\sigma}_{s's})^a}{2m}
    %\qquad
    %:
    %\qquad
    %a \in \{1,2,3\}
    \\
    \llbracket
    \gamma^-
    \rrbracket
    &=
    \frac{1}{m}
    \left\{
      \left(
      E_0
      +
      \frac{\bm{P}^2}{2m}
      -
      \frac{\bd^2}{8m}
      \right)
      \delta_{s's}
      -
      \frac{i (\bd\times\bm{\sigma}_{s's})\cdot\bm{P}}{2m}
      \right\}
    \\
    \llbracket
    \sigma^{ab}
    \rrbracket
    &=
    \sqrt{\frac{2E_0}{m}}
    \epsilon_{abc}
    (\sigma_b)_{s's}
    %\qquad
    %:
    %\qquad
    %a,b \in \{1,2,3\}
    \\
    \llbracket
    \sigma^{+a}
    \rrbracket
    &=
    0
    %\qquad
    %:
    %\qquad
    %a \in \{1,2,3\}
    \\
    \llbracket
    \sigma^{-a}
    \rrbracket
    &=
    \sqrt{\frac{2E_0}{m}}
    \left(
    \frac{\epsilon_{abc}(\sigma_b)_{s's}P_c}{m}
    -
    \frac{i\dl_a}{2m}
    \right)
    %\qquad
    %:
    %\qquad
    %a \in \{1,2,3\}
    \\
    \llbracket
    \sigma^{+-}
    \rrbracket
    &=
    0
    \,,
  \end{split}
\end{align}
where $a,b \in \{1,2,3\}$.

% Levy-Leblond spinors
\section{L\'evy-Leblond spinors}
\label{sec:levyleblond}

L\'evy-Leblond spinors~\cite{Levy-Leblond:1967eic}
are the standard set of spinors used in literature
on Galilean quantum mechanics and field
theory~\cite{Hagen:1970cj,Hurley:1971nz,omote:gal,deMontigny:2003gdw,Santos:2004pq,deMontigny:2006,Niederle:2007xp,Huegele:2011,Wilkes:2020}.
This appendix briefly introduces them
and explains why I am not using them in this work.
The main issue is that they don't have the expected transformation
properties under Galilean parity reversal,
which would limit the number of true tensor
(as opposed to pseudotensor)
structures that could be added to form factor breakdowns.
For instance, this means L\'evy-Leblond spinors
cannot encode an anomalous magnetic moment, nor internal stresses.

The L\'evy-Leblond equation can be found quickly
(albeit in a non-standard form)
using the tau-matrix formalism of Sec.~\ref{sec:spinor}.
The key trick is to note that
if $\psi_s(p)$ transforms under the complex defining representation
of $\mathrm{USp}(2,2,\mathbb{C})$,
then so does $p^\mu \tau_\mu \eta \psi_s(p)$.
Accordingly, the following momentum-space equation is covariant
under the connected part of the Galilei group:
\begin{align}
  \label{eqn:ll:E0}
  p^\mu \tau_\mu
  \eta \psi_s(p)
  =
  \big(
  m \tau_+
  +
  E \tau_-
  -
  \bm{p}\cdot\bm{\tau}
  \big)
  \eta \psi_s(p)
  =
  \sqrt{2mE_0}
  \psi_s(p)
  \,,
\end{align}
where the $\tau_\mu$ matrices are defined in Eq.~(\ref{eqn:vm}).
This equation is not, however, invariant under Galilean parity
reversal unless $E_0=0$.
To see this, parity reversal is effected
by acting on on Eq.~(\ref{eqn:ll:E0})
with $\tau_4$ (see Sec.~\ref{sec:parity}).
Since $\tau_4\eta = -\eta\tau_4$:
\begin{align}
  p^\mu \tau_\mu
  \eta \psi_s(p)
  =
  \big(
  m \tau_+
  +
  E \tau_-
  +
  \bm{p}\cdot\bm{\tau}
  \big)
  \eta
  \big( \tau_4 \psi_s(p) \big)
  =
  -
  \sqrt{2mE_0}
  \big( \tau_4 \psi_s(p) \big)
  \,,
\end{align}
which is not equivalent to Eq.~(\ref{eqn:ll:E0}) except when $E_0=0$.
Setting $E_0=0$ gives:
\begin{align}
  \label{eqn:ll}
  p^\mu \tau_\mu
  \eta \psi_s(p)
  =
  \big(
  m \tau_+
  +
  E \tau_-
  -
  \bm{p}\cdot\bm{\tau}
  \big)
  \eta \psi_s(p)
  =
  0
  \,,
\end{align}
which is in effect the L\'evy-Leblond equation~\cite{Levy-Leblond:1967eic}.
To be sure, this is not the standard form of the equation,
but it can be proved that there is (up to equivalences)
a unique Galilei-covariant wave equation for
four-component spinors~\cite{Huegele:2011}.

Now, even though Eq.~(\ref{eqn:ll}) itself is covariant under parity reversal,
its solutions have unexpected parity transformation properties.
Since $\tau_4\eta = -\eta\tau_4$,
$\psi_s(p)$ and $\eta\psi_s(p)$ have opposite parity,
and consequently the bilinear
$\psi_s^\dagger(p) \eta \psi_s(p)$---the equivalent
of the Dirac scalar $\bar{u}(p,s)u(p,s)$---is a pseudoscalar.
This means $\bar{u}u$-like terms cannot appear in form factor breakdowns
if the current in question is a true tensor.

I will presently explore this issue more concretely
in terms of the standard form of the L\'evy-Leblond equation.

%%%%%%%%%%%%%%%%%%%%%%%%%%%%%%%%%%%%%%%%%%%%%%%%%%%%%%%%%%%%%%%%%%%%%%%%%%%%%%%%

\subsection*{Standard form of the L\'evy-Leblond equation}

The L\'evy-Leblond equation can be written in its standard form using the
unitary transformation matrix:
\begin{align}
  \label{eqn:ll:map}
  L
  =
  \frac{1}{2}
  \begin{bmatrix}
    1+i & 1+i
    \\
    i-1 & 1-i
  \end{bmatrix}
  \,,
\end{align}
and defining the quantities:
\begin{align}
  \label{eqn:ll:gamma}
  \begin{split}
    u_{\levy}(p,s)
    &=
    L
    \psi_s(p)
    \\
    \gamma_{\levy}^\mu
    &=
    L \tau^\mu \eta L^{-1}
    \,,
  \end{split}
\end{align}
for which the explicit gamma matrices are:
\begin{align}
  &
  \gamma_{\levy}^0
  =
  \begin{bmatrix*}[r]
    0 & -i \\
    i & 0
  \end{bmatrix*}
  \qquad
  \gamma_{\levy}^a
  =
  \begin{bmatrix}
    i\sigma_a & 0 \\
    0 & -i\sigma_a
  \end{bmatrix}
  \qquad
  \gamma_{\levy}^4
  =
  \begin{bmatrix}
    0 & i \\
    i & 0
  \end{bmatrix}
  \,.
\end{align}
Note that the gamma matrices defined in this way obey
the Clifford algebra relation
$\{\gamma_{\levy}^\mu,\gamma_{\levy}^\nu\} = 2 g^{\mu\nu}$.
Plugging these definitions into Eq.~(\ref{eqn:ll}) gives:
\begin{align}
  \label{eqn:ll:std}
  \slashed{p}_{\levy}
  u_{\levy}(p,s)
  =
  -
  i
  \begin{bmatrix}
    \bm{p}\cdot\bm{\sigma} &  \sqrt{2} m \\
    -\sqrt{2} E & - \bm{p}\cdot\bm{\sigma}
  \end{bmatrix}
  u_{\levy}(p,s)
  =
  0
  \,.
\end{align}
This is the standard form of the L\'evy-Leblond equation,
given for instance
(up to a difference of sign convention in the metric)
by Refs.~\cite{omote:gal,deMontigny:2003gdw,Santos:2004pq}.

Defining two-component sub-spinors:
\begin{align}
  \phi_\uparrow
  =
  \begin{bmatrix}
    1 \\ 0
  \end{bmatrix}
  \qquad
  \phi_\downarrow
  =
  \begin{bmatrix}
    0 \\ 1
  \end{bmatrix}
  \,,
\end{align}
solutions to Eq.~(\ref{eqn:ll:std}) can be written:
\begin{align}
  \label{eqn:levysol}
  u_{\levy}(p,s)
  =
  \frac{1}{\sqrt{2}}
  \begin{bmatrix}
    \phi_s
    \\
    -
    \frac{\bm{p}\cdot\bm{\sigma}}{\sqrt{2}m}
    \phi_s
  \end{bmatrix}
  \,.
\end{align}

The parity problem arises when building bilinear forms in
$u_{\levy}$ and its Hermitian conjugate.
Form factor breakdowns like Eq.~(\ref{eqn:emt:half})
need to be true Galilei tensors, rather than pseudotensors.
However,
$\bar{u}_{\levy}(p',s') u_{\levy}(p,s)$
is a pseudoscalar rather than a true scalar,
which limits the number of possible tensors that can be built.
For instance, form factors akin to $D_q(\bd^2)$ and $\bar{c}_q(\bd^2)$
could never appear in a breakdown with L\'evy-Leblond spinors.

To see this, the conjugate spinor is as usual defined
via $\bar{u}_{\levy} = u^\dagger_{\levy} \gamma^0_{\levy}$.
By the change of basis in Eq.~(\ref{eqn:ll:map}), this is equivalent to making
$\bar{\psi}_s(p) = \psi^\dagger_s(p)\eta$ the conjugate of $\psi_s(p)$
in Eq.~(\ref{eqn:ll}).
This makes sense, since under the path-connected part of the Galilei group,
$\bar{\psi}_s(p) \mapsto \bar{\psi}_s(p) M^{-1}$,
making
$\bar{u}_{\levy}(p',s') u_{\levy}(p,s)$
transform like a scalar under this subgroup.
However, this quantity is parity-odd, because $\eta\tau_4\eta = -\tau_4$.
Thus
$\bar{u}_{\levy}(p',s') u_{\levy}(p,s)$
is a pseudoscalar.
Explicit calculation even shows:
\begin{align}
  \bar{u}_{\levy}(p',s') u_{\levy}(p,s)
  =
  -
  \frac{i\bd\cdot\bm{\sigma}}{2m}
  \,,
\end{align}
which is clearly parity-odd,
and would indicate an intrinsic electric dipole moment
if it appeared in an electromagnetic form factor breakdown.

In fact, for electromagnetic form factors,
the most general $(\text{Galilei}+\text{parity})$-covariant expression
that can be built using L\'evy-Leblond spinors is:
\begin{align}
  \langle \bm{p}', s' |
  \hat{J}^\mu(0)
  | \bm{p}, s \rangle
  =
  \bar{u}_{\levy}(p',s')
  \gamma_{\levy}^\mu
  u_{\levy}(p,s)
  F(\bd^2)
  \,.
\end{align}
Explicitly evaluating the $(+,1,2,3)$ components for the charge density
and current gives:
\begin{align}
  \begin{split}
    \langle \bm{p}', s' |
    \hat{J}^\mu(0)
    | \bm{p}, s \rangle
    &=
    \delta_{s's}
    F(\bd^2)
    \,,
    \\
    \langle \bm{p}', s' |
    \hat{J}^a(0)
    | \bm{p}, s \rangle
    &=
    \left(
    \frac{P^a}{m}
    \delta_{s's}
    -
    \frac{i (\bd\times\bm{\sigma}_{s's})^a }{2m}
    \right)
    F(\bd^2)
    \,,
  \end{split}
\end{align}
which cannot encode an anomalous magnetic dipole moment.

%%%%%%%%%%%%%%%%%%%%%%%%%%%%%%%%%%%%%%%%%%%%%%%%%%%%%%%%%%%%%%%%%%%%%%%%%%%%%%%%

\bibliography{references.bib}

\end{document}